\documentclass[a4paper,fleqn,usenatbib]{mnras}

\usepackage[T1]{fontenc}
\usepackage{ae,aecompl}
\usepackage{pdflscape}

\usepackage{graphicx}	
\usepackage{amsmath}	
\usepackage{amssymb}	
\newcommand{\Hb}{H$\beta$}
\newcommand{\Ha}{H$\alpha$}

\newcommand{\SHaObs}{{\it S(H$\rm\alpha,OBS$)}}
\newcommand{\SHbObs}{{\it S(H$\rm\beta,OBS$)}}
\newcommand{\SHa}{{\it S(\rm H$\alpha$)}}
\newcommand{\SHb}{{\it S(\rm H$\beta$)}}
\newcommand{\sbunits}{ergs~s$^{-1}$~cm$^{-2}$~sr$^{-1}$}
\newcommand{\kms}{km~s{$^{-1}$}}
\newcommand{\cmq}{cm$^{-3}$}

\newcommand{\pers}{s$^{-1}$}

\newcommand{\tc}{{$\theta^1$~Ori~C}}
\newcommand{\ta}{{$\theta^2$~Ori~A}}
\newcommand{\hii}{H{\sc ii}}
\newcommand{\cii}{[\ion{C}{\sc ii}]}
\newcommand{\sii}{[\ion{S}{\sc ii}]}
\newcommand{\siii}{[\ion{S}{\sc iii}]}
\newcommand{\oi}{[\ion{O}{\sc i}]}
\newcommand{\oii}{[\ion{O}{\sc ii}]}
\newcommand{\oiii}{[\ion{O}{\sc iii}]}
\newcommand{\nii}{[\ion{N}{\sc ii}]}

\newcommand{\Ne}{{\it n$\rm_{e}$}}
\newcommand{\Te}{{\it T$\rm_{e}$}}
\newcommand{\Th}{{\it T$\rm_{e}$(H$^{+}$)}}
\newcommand{\Tn}{{\it T$\rm_{e}$([N~II])}}
\newcommand{\cHb}{{\it c$_{H\beta}$}}
\newcommand{\Cvla}{{\it c$\rm_{VLA}$}}
\newcommand{\Cmuse}{{\it c$_{MUSE}$}}

\newcommand{\Hplus}{H$^{+}$}
\newcommand{\Heplus}{He$^{+}$}
\newcommand{\ozone}{He$^{+}$+H$^{+}$}
\newcommand{\hr}{Huygens Region}
\newcommand{\bb}{Bright Bar}
\newcommand{\nzone}{He$\rm ^{o}$+H$^{+}$}
\newcommand{\vrad}{{\it V$\rm _{r}$}}   

\newcommand{\Vblue}{{\it V$\rm_{Blue}$}}

\newcommand{\Vexp}{{\it V$\rm_{exp}$}}
\newcommand{\Vpdr}{{\it V$\rm_{PDR}$}}

    \title[Structure and  Conditions in Orion]{{\it Structure and Physical Conditions in the Huygens Region of the Orion Nebula}}

\author[C. R. O'Dell et al.]{
C. R. O'Dell,$^{1}$\thanks{E-mail: cr.odell@vanderbilt.edu (CRO)}
G. J. Ferland,$^{2}$
and M. Peimbert$^{3}$\\
$^{1}$Department of Physics and Astronomy, Vanderbilt University, Box 1807-B, Nashville, TN 37235 USA\\
$^{2}$Department of Physics and Astronomy, University of Kentucky, Lexington, KY 40506 USA\\
$^{3}$Instituto de Astronomia, Universidad Nacional Aut\'onoma de M\'exico, Apdo, Postal 70-264, 04510 M\'exico D. F., M\'exico}

\date{Accepted 2016 October 19. Received 2016 October 17; in original form 2016 May 27}

\pubyear{2016}

\begin{document}
\label{firstpage}
\pagerange{\pageref{firstpage}--\pageref{lastpage}}
\maketitle

\begin{abstract}


HST images, MUSE maps of emission-lines, and an atlas of high velocity resolution emission-line spectra have been used to establish for the first time correlations of the electron temperature, electron density, radial velocity, turbulence, and orientation within the main ionization front of the nebula. 

From the study of the combined properties of multiple features, it is established that variations in the radial velocity are primarily caused by the photo-evaporating ionization front being viewed at different angles.

There is a progressive increase of the electron temperature and density with decreasing distance from the dominant ionizing star \tc. 
The product of these characteristics (\Ne $\times$ \Te) is the most relevant parameter in modeling a blister-type nebula like the Huygens Region, where this quantity should vary with the surface brightness in \Ha. 

Several lines of evidence indicate that small-scale structure and turbulence exists down to the level of our resolution of a few arcseconds.


Although photo-evaporative flow must contribute at some level to the well-known non-thermal broadening of the emission lines, comparison of quantitative predictions with the observed optical line widths
 indicate that it is not the major additive broadening component.


Derivation of \Te\ values for \Hplus\ from radio+optical and optical-only ionized hydrogen emission showed that this temperature is close to that derived
from \nii\ and that the transition from the well-known flat extinction curve that applies in the \hr\ to a more normal steep extinction curve occurs immediately outside of the Bright Bar feature of the nebula.

\end{abstract}



\section{introduction}
\label{sec:intro}
The basic model of the \hr\ of the Orion Nebula is that of an irregular concave blister of ionized gas  located on the observer's 
side of a giant molecular cloud (a summary is found in \citep{ode01}). \citet{vdw89} established through observations of the HI 21 cm line in absorption 
that there was a foreground Veil of primarily neutral material. The Veil has been the subject of several
additional studies using radio and optical absorption lines \citep{vdw13,abel16}. These studies refine the properties of the Veil, although its distance from 
the dominant ionizing star (\tc) remains uncertain. The study of the correlation of HI absorption and reddening by \citet{ode92} established
that the extinction of the nebula and the Trapezium stars arise within the foreground Veil. Within the space between the concave blister of ionized gas and the Veil, there is a molecular cloud called Orion-South (sometimes Orion-S) which is ionized on the observer's side, but molecular absorption lines
indicate that it lies in front of another layer of ionized gas. Within the volume between \tc\ and the Veil there is a blue-shifted layer of ionized gas (the Blue Layer) whose origin and properties are uncertain.

The gas density within the blister of gas near the Main Ionization Front (MIF) decreases rapidly with increasing distance from the MIF (towards the 
observer). This ionized gas is stratified in ionization, each layer having an easily observed strong emission line. \oi\ emission arises from immediately at the MIF.
Closer to \tc\ there is a layer of \nzone\ giving rise to \nii\ emission. This layer is about 1\arcsec\ thick \citep{ode01}. That layer gives
way to the higher ionization \ozone\ layer , which is about 30\arcsec\ thick \citep{ode01}, and produces strong \oiii\ emission. The Balmer lines (\Ha\ and \Hb )
arise from both layers. \tc\ is too cool to produce a higher ionization layer.

The brightest portion of the Orion Nebula (M~42, NGC 1976) is commonly designated as the \hr, as it was first depicted in print by Christiaan Huygens in the mid 17th century \citep{gin}. The \hr\ has been the subject of many studies at essentially all available energies, but defies a definitive resolution of its structure and physical conditions. Recent reviews \citep{mu08,ode08} describe in more detail our knowledge of
the nebula and its associated young stars.

Although the basic model of the nebula seems to be understood, there are many unresolved issues. 
What is the cause of the variations in radial velocity across the face of the nebula? Is there evidence for small-scale variations in the electron temperature?  What is the cause of the known electron density and electron temperature variations? What is the source of the unexpected width of the emission lines?
This list could be continued, but the important bottom-line is that the \hr\ of the Orion Nebula is our best 
opportunity for studying other H~II regions, both in our Galaxy and in extragalactic systems; if we cannot understand the processes 
occurring in this object, conclusions about other similar but more poorly characterized objects are rendered correspondingly uncertain.


In this article we report on an investigation of the physical conditions within the \hr\ and how these relate to features near the MIF. 
Such an investigation was made possible through the existence of an atlas of high velocity resolution \citep{gar08} (hereafter the Spectral Atlas), numerous Hubble Space Telescope images, and a recent moderate resolution spectroscopic map \citep{wei15}. In Section \ref{sec:data} we describe the data used and their processing.
In Section~\ref{sec:Vcorr} we describe the basic premise in interpretation of the radial velocities. In Section~\ref{sec:overview} we present the study of individual features and regions and how this reveals their structure and physical conditions. In Section~\ref{sec:outflows} we explore 
the variations of \Ne, \Te, and structure in the eastern region of the Orion-South Cloud, which contains multiple high velocity jets and shocks.
In Section~\ref{sec:fwhm} we discuss the meaning of the observed large values of the Full Width at Half Maximum (FWHM). A summary of the results of relating \Te\ to other nebular characteristics is in Section~\ref{sec:NeTe}. What the lack of a back-scattered component tells us about the 3-D structure of some samples is presented in Section~\ref{sec:BackScat}.
The results of an attempt to determine \Te\ for \Hplus\ is given in Appendix \ref{sec:IntroExt}. A summary of this study is presented in Section \ref{sec:conclusions}.

\section{The data sets employed}
\label{sec:data}

\subsection{Imaging Data}
\label{sec:ImagingDat}
The \hr\  was a natural target for one of the first observations with the MUSE moderate resolution spectrograph \citep{wei15}.
We draw heavily upon this data-set, which generously is in the public domain.The images were reprocessed, using the original images, as described in our paper on fine scale temperature fluctuations in the \hr\ (William Henney, et al., in preparation). Although the fiber
optics defining the sampling has a size of 0\farcs2 over the 353\farcs2$\times$295\farcs2 Field of View (FOV), the images created are about 1\arcsec\ FWHM. A shortcoming of this data set is that only
a single exposure time was used for each pointing, so that the signal to noise (S/N) ratio decreases as one utilizes the weaker lines. For our
study, we employed only strongest lines (486.1 nm, \Hb; 500.7 nm, \oiii; 656.3 nm, \Ha; 658.3 nm, \nii). The derivation of \Te\ was done
using both the strong 658.3 nm nebular line and the weaker 575.5 nm auroral line, while \Ne\ was determined from the moderately 
strong red \sii\ doublet (671.6 nm and 673.1 nm). Throughout this paper we will use the \Te\ and \Ne\ values determined by Henney, et al.  Unfortunately, the low velocity resolution (107 \kms) of the MUSE spectra preclude
study of any but the highest radial velocity (\vrad) features.

Although monochromatic images can be made of the full MUSE field of view, the superior resolution
HST images were employed where possible. The original mosaic of images using the lower resolution WFPC2 camera images \citep{ode96} were frequently 
employed, although in a few cases the better sampled PC camera images were available \citep{ode97}. The HST images (about 0\farcs08)were calibrated into flux
units using the calibration constants determined by comparing regions within these images with flux calibrated ground-based 
telescope observations \citep{ode09a}. For our purposes it was sufficient to ignore the low level of contamination of the HST emission-line filters
by the underlying continuum and for contaminating lines falling in the wings of the filters. 

\subsection{Spectra}
\label{sec:spectra}

The high velocity resolution spectra  we used are from the Spectral Atlas compiled by \citet{gar08}. These have a velocity resolution of about 8--10 \kms.
The Spectral Atlas was compiled from a series of north-south oriented slit spectra made with a spacing of 2\arcsec\ under conditions of astronomical seeing better than that. This means that the Spectral Atlas has a spatial resolution of about 2\arcsec. Although the Spectral Atlas includes 
incomplete mapping in additional lines, we employed only the fully mapped \oiii\ 500.7 nm and \nii\ 658.3 nm lines, which were 
calibrated to about 2 \kms\ accuracy.

These high S/N spectra of \oiii\ and \nii\ are intrinsically narrow and can be de-convolved into multiple components. This was not 
possible with the ubiquitous \Ha\ and \Hb\ lines because of their large thermal width. As done in earlier studies of the \hr\ de-convolution
was done by using the IRAF  \footnote{IRAF is distributed by the National Optical
Astronomy Observatories, which is operated by the Association of
Universities for Research in Astronomy, Inc.\ under cooperative
agreement with the National Science foundation.} task "splot". In addition to the strongest component (produced on the observer's side 
of the MIF) there is almost always a red wing of about 10\%\ the strength of the MIF component (produced by backscattering from the 
dust in the compressed neutral zone beyond the MIF), and often a weaker blue component. The \vrad\ of the MIF component
could usually be determined with an accuracy of 1 \kms\ or better, while the accuracy of the red and blue radial velocities are about 3 \kms, the exact value depending on the flux relative to the MIF and the velocity separation from that same strong component. The 
origin of the red component as backscattered light is by now well established  \citep{ode92,hen94,ode01,abel16} as it is fully consistent with the blister model for the MIF and the photo-evaporation velocity shift between the emitting and scattering layers produces a doubling of the observed velocity separation. One would expect that the relative strength of the backscattered component to be weak compared with the MIF component when seeing a highly tilted portion of the MIF. The few absences of a backscattered component are discussed in Section~\ref{sec:BackScat}. The more difficult to observe blue component is best described as a layer of ionized gas lying between \tc\ and the foreground Veil \citep{abel16}.

\section{Interpretation of variations of the MIF Radial Velocities}
\label{sec:Vcorr}

The radial velocities of the MIF components are particular important in studying the inclination of the local MIF. For a photo-evaporating
blister nebula gas is ionized at the MIF, then flows away from the Photo Dissociation Region (PDR). The \nii\ emission arises from
closer to the MIF and has a smaller (blueshifted) velocity relative to the PDR than the \oiii\ emission, which is from gas that has been
accelerated more as it reaches a larger distance from the MIF. Observations of \cii\ \citep{goi15} at a spatial resolution of about 11\farcs4 indicate
that the \vrad\ of the PDR (which gives rise to the \cii\ emission) is constant at 27.5$\pm$1.5 \kms (LSR) across the \hr\ and are indistinguishably the same as the average of the lower spatial resolution CO velocities (27.3$\pm$0.3 \kms\ LSR (derived from Table 3.3 VII of \citet{gou82}. If the MIF
was at a fixed angle with respect to the sky, then one would expect \vrad\ for \oiii\ and \nii\ to be constant, with \oiii\ more blueshifted than \nii. However the observed \vrad\ varies for both lines across the \hr. For a face-on MIF the observed \vrad\ would be \Vpdr\ - \Vexp, where  \Vpdr\ is the velocity of the PDR and \Vexp\ is the photo-expansion velocity. In the case of 
the MIF being viewed edge-on, the observed \vrad\ would be \Vpdr, since the \Vexp\ component will then be perpendicular to the line of sight. Throughout
this paper we will interpret variations in the observed \vrad\ as due to variations in the tilt of the MIF.

\section{Areas and Features Discussed in Detail in this Paper}
\label{sec:overview}

\begin{figure*}
	\includegraphics
	[width=7in]
	{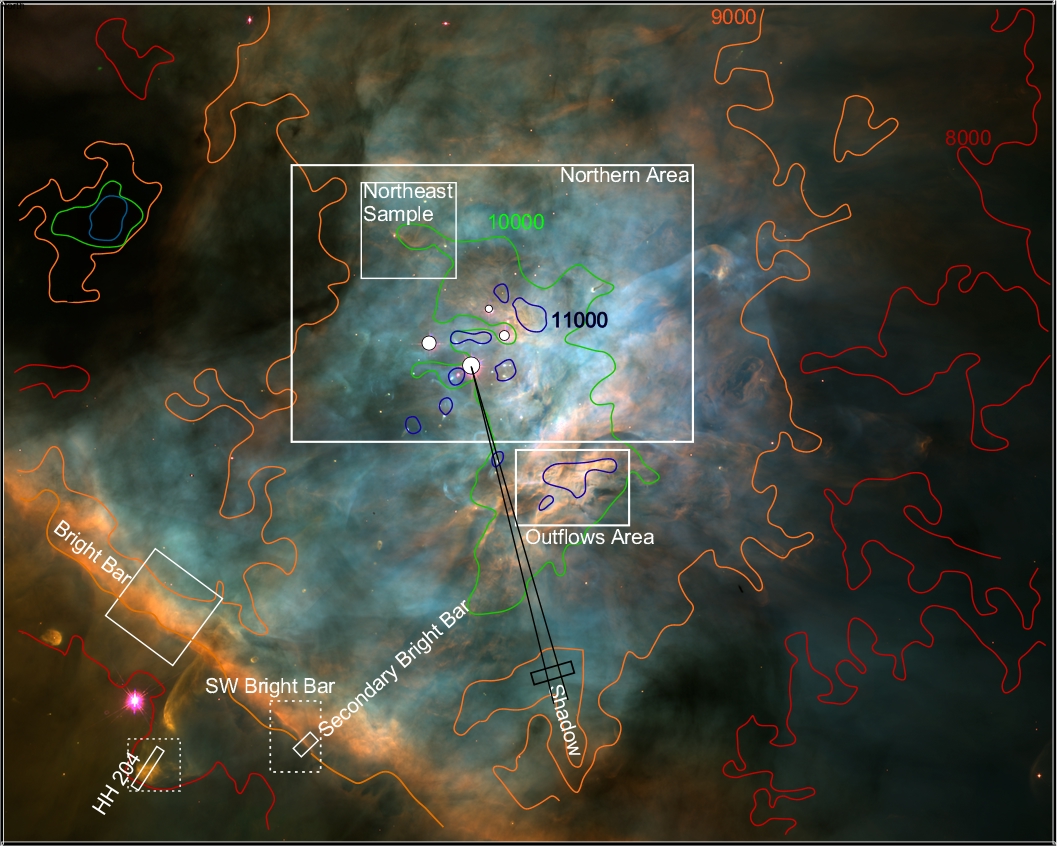}
    \caption{
    This 204\farcs2$\times$243\farcs6 FOV image of the \hr\ of the Orion Nebula is taken from a mosaic of WFPC2 HST images \citep{ode96},
    where the colour coding is Blue-F502N (\oiii), Green-F656N (\Ha), and Red-F658N (\nii) and is slightly smaller than the MUSE \citep{wei15} FOV. The colour contours are the \nii\ \Te\ values derived from our processing of the MUSE data (W. J. Henney-in preparation). Rectangular boxes 
    indicate features discussed in various sections of the text. The two lines extending from the brightest star (\tc, 5:35:16.46 -5:23:22.9) in the Trapezium 
    grouping indicate the alignment with obscuring material that defines the ionization shadow discussed in Section \ref{sec:shadow}. One dashed box indicates the fine-structure portion of the \bb\ that we designate as the SW \bb. The other dashed box indicates the region around HH~204 studied by \citet{nun12}.}
     \label{fig:HSTfov}
\end{figure*}

We have sampled multiple regions within the \hr, as illustrated in Figure \ref{fig:HSTfov}. They were selected to cover a variety of characteristics. HH~204 (Section \ref{sec:HH204}) is a well defined shock structure seen to advantage against the much lower 
surface brightness of the nebula outside of the \bb. 
Linear features are studied in the region of the \bb\ (Section~\ref{sec:BrightBar}), the SW \bb\ (Section~\ref{sec:narrow}), and within an ionization shadow (Section~\ref{sec:shadow}), with the former two represented regions where the MIF is viewed almost edge-on. Within the Northern Region there is a large high \Te\ arcuate structure (Section \ref{sec:NE}) that is a structure within the MIF in spite of its resemblance to a bow shock. The complex Northern Region is discussed in Section~\ref{sec:NorthernArea} where the effect of local variations in structure of the MIF are illuminated. Finally, the regions of multiple high velocity outfows are discussed in Section~\ref{sec:outflows}.


\subsection{HH 204}
\label{sec:HH204}
\begin{figure*}
\includegraphics
[width=6in]
{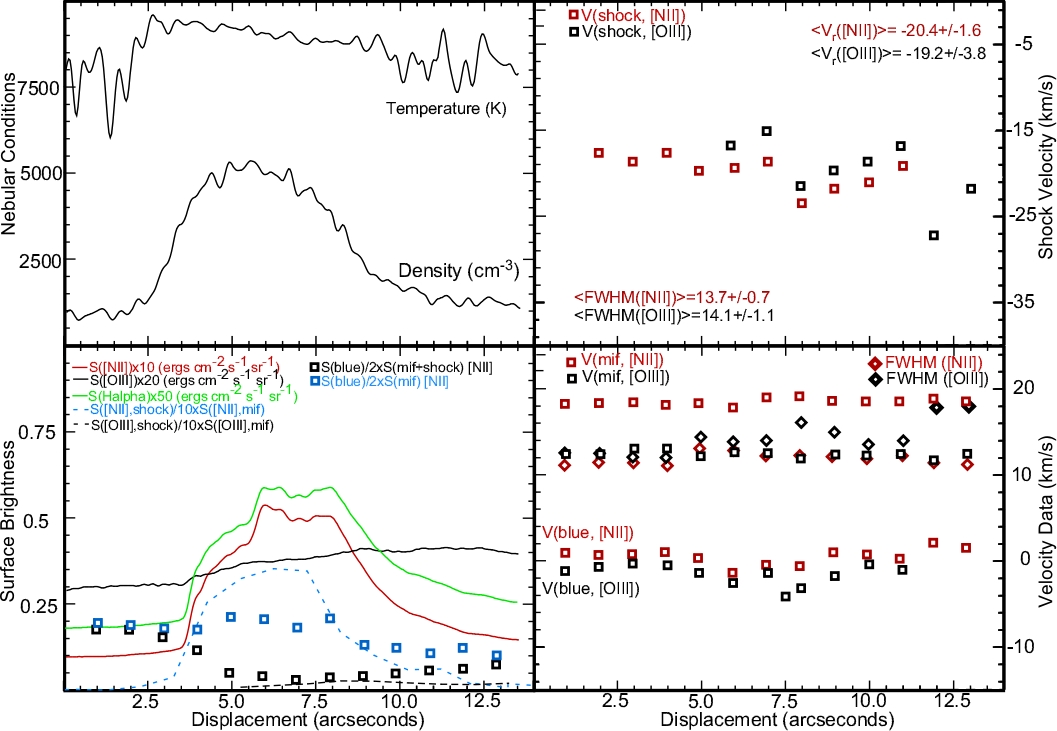}
\caption{This figure shows various quantitative results derived for the sample marked HH~204 in Figure~\ref{fig:HSTfov}. Given the large amount of data and the need to compress its presentation, in some cases the data has been scaled. Increasing displacement corresponds to displacement along the shock towards PA~=~327\degr.}\label{fig:HH204}
\end{figure*}

There is a rich literature of observations of the object now designated as HH~204. It has been imaged at highest resolution with the PC camera of the HST \citep{ode97}, its motion in the plane of the sky has been determined (Doi, O'Dell, \&\ Hartigan, 2002), and its \vrad\ mapped at high spectral resolution (Doi, O'Dell, \&\ Hartigan, 2004).  \citet{mesa08} obtained a long-slit spectrum along the axis of symmetry of the shock and a region of 15\arcsec\ $\times$~15\arcsec\  at the apex was mapped with 1\arcsec,\ pixels with integral field spectroscopy of 3.5 \AA\ resolution \citep{nun12}. In spite of this variety of studies, more can be done, especially with the use of high velocity resolution data.

The HH~204 shock is driven by a high velocity jet originating near the eastern side of the Orion-South molecular cloud.  The most recent discussion of the probable origin is found in \citet{ode15}. The shock moves toward the observer with a spatial velocity of 103 \kms\ at an angle of about 27\degr\ out of the plane of the sky \citep{hen07}.  It is unclear if it is formed as the jet strikes the foreground Veil \citep{ode97} or the 
raised escarpment that produces the \bb\ \citep{doi04,vdw13}. It is pointed away from \tc\ and its structure is consistent
with a shock that is photoionized from within, that is, the material within the parabolic envelope of the shock is illuminated and ionized 
by \tc.  

We created a sample along the slit spectrum of \citet{mesa08} and within the area studied by \citet{nun12}. Our sample is 2\farcs4 wide and 
13\farcs5 long. The physical conditions were sampled from the \sii -derived \Ne\ and \nii -derived \Te\  approximately 1\arcsec\ resolution maps using the MUSE data. The results are shown in the top-left panel of Figure \ref{fig:HH204} and are similar the results presented in
Figure 7 of \citet{nun12}. It should be noted that the zero-point of our samples is at -3\farcs2 on the \citet{nun12} plot.  
Higher spatial resolution (about 0\farcs1) samples of the surface brightness were made using the calibrated WFPC2 images of \nii, \Ha, and \oiii. These are shown
in the lower left panels of Figure \ref{fig:HH204}. It should be noted that the WFPC2 filters would have transmitted both the MIF and the highly
blueshifted components. 

The same sampling
was done using the \nii\ and \oiii\ lines from the Spectral Atlas. In this case the sampling distance was 
0\farcs5 (however, recall that the spacing of the Spectral Atlas  North-South slits is 2\arcsec\ while the sampling in each line is 0\farcs53 and characteristic seeing was about 2\arcsec) and each of these pseudo-spectra samples of 2\farcs4$\times$0\farcs5 was de-convolved using "splot".  As presented in Section \ref{sec:data}, the \hr\ spectra usually de-convolve into the MIF, Red (backscattered), and Blue (an ionized layer). In addition to these three components, we also saw a much more blueshifted component arising in shocked gas.  The radial velocities for the shocked components are shown in the top-right panel of Figure \ref{fig:HH204}, in addition to the average of the FWHM of these shock components. The average FWHM values are only slightly larger than those for the 
individual samples of the MIF components, as shown in the lower-right panel.  The shocked material velocities are well blueshifted with
respect to the MIF velocities shown in that same lower-right panel.  The blue component velocities in the lower-right panel are at values
similar to much of the rest of the nebula (0.9$\pm$2.8 \kms\ for \oiii\ and 1.8$\pm$1.9 \kms\ for \nii, \citep{abel16}). 

In this object and the other subsequent objects addressed we have adopted a procedure to assure that the slit spectra samples align
with the images used to produce \Ne , \Te , and emission-line profiles. We created from the spectra the total surface brightness at each slit. The
total surface brightness includes all of the components of the line. The sequence of slits was shifted to provide agreement of the image profile (the filters are wide enough to include emission from all the velocity components) and the results from the spectra.

The abrupt drop in S(\nii) and S(\Ha) at 3\farcs5 indicates that there is an ionization boundary associated with the shock, as argued in \citet{nun12}.
At that same point we see S(\oiii), drop slightly from an enhanced level to what must be the nebular background level. This supports the argument that the envelope of the shock is photoionized from the inside by \tc.  The density (rising to 4000 \cmq\ over ambient) agrees in
location with the S(\nii) and S(\Ha) increase over 3\arcsec --9\arcsec. 

We also give the ratio of surface brightness of the shocked over the MIF components (the dashed lines) in the lower-left panel of Figure~\ref{fig:HH204}.
We see there that the \nii\ shock component reaches 3.6 times the strength of the MIF component and that the \oiii\ shock component
reaches only 0.3 times the strength of the MIF component. The WFPC2 image of \oiii\ has a much higher S/N ratio and the fact the 
the total surface brightness in \oiii\ determined from that data extends to the shock boundary defined by the drop in S(\Ha) and S(\nii)
means that shocked \oiii\ material fills the envelope of the shock all of the way to the ionization boundary. In addition, the dominance
of shocked \nii\ beyond the ionization boundary indicates that it is shocked gas producing the density jump.

Not only do we confirm the model for HH~204 presented by \citet{nun12}, we also confirm that there is a narrow zone of high \Te\ material just 
outside of what one would ordinarily call the ionization boundary.  The biggest step in the present study is the use of high velocity spectra
to clearly link the \Te\ and \Ne\ enhancements to the shocked material.

The blue component (whose velocities are well separated from those of the MIF and the shocks) is unusually strong. In the samples of 
\citet{abel16} they found that the Blue/MIF surface brightness ratio was about 0.08$\pm$0.04.  In the lower-left panel of Figure~\ref{fig:HH204} we 
see that the observed ratio varies greatly, depending on whether one is taking as the denominator the sum of the MIF and Shock components
or the MIF component alone. It is most likely that the ratio using only the MIF component is relevant and its ratio is nearly constant along the 
sample (an average of 0.31$\pm$0.09). This indicates that the Blue Layer is independent of the HH~204 shock.  However, it is probably
significant that the ratio of the Blue/MIF components in this sample (all outside of the \bb) is some four times larger than the central
nebula samples of \citet{abel16} and rises in our HH~204 sample with increasing distance from the \bb.  As we shall see in Section \ref{sec:Difr}, there are arguments for abrupt changes in the foreground material as one 
goes beyond the \bb.

\subsection{The \bb}
\label{sec:BrightBar}

The feature known as the \bb\ is well studied and beyond the fact that it is basically an escarpment viewed nearly from along the cliff face it is not well understood, both at a fine-scale (Is it a simple curved surface or an undulating surface viewed almost edge-on?) and at a large scale (How does one 
produce a structure in the MIF that is so linear?).  Listings of the arguably best earlier studies of the ionized side (facing \tc) and neutral side (mostly molecules) 
are presented in \citep{wal00,ode08}. More recently there have been additional valuable observational studies \citep{vdw09,mesa11} and models \citep{hen05b,pel09,shaw09,asc11}. \citet{gar07} present evidence for several other bar structures, but none are as obvious as the \bb.
In this study we have sought to bring together the recently available MUSE material, the high velocity resolution Spectral Observations material, and HST optical imaging to produce a more accurate picture of what it is. 

In order to obtain high S/N data, we have grouped the data within the \bb\ sample box (Figure~\ref{fig:HSTfov}
into narrow samples parallel (PA = 53.4\degr) to the \bb. Each sample was 24\arcsec\ long and the artificial spectra was created similar to the study of HH~204 and have a spacing of 1\arcsec. There is a certain amount of averaging along each sample because of the fine-scale structure of the \bb. This reduces the inherently 
high resolution of the HST images but probably does not change the 2\arcsec\ resolution inherent in the Spectral Atlas.
The MUSE data of about 1\arcsec\ resolution was probably degraded slightly by this process, but not by much. 

\begin{figure*}
	\includegraphics
	[width=7in]
	{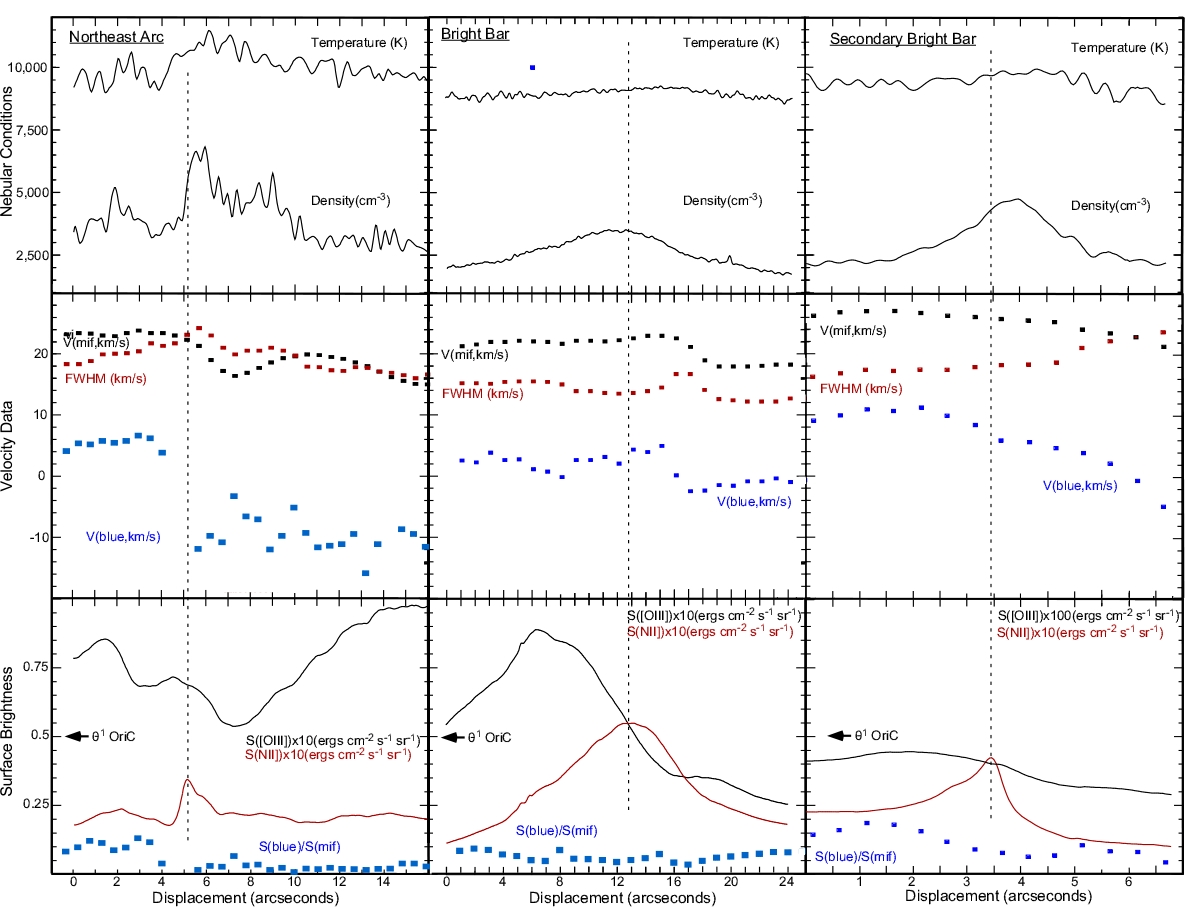}
    \caption{
  The results of profiles taken across different samples are shown as a function of increasing distance from the end of the sample closest 
  to \tc. 
 The top panels show the profile of \Te\ and \Ne. The middle 
  panels show the results of using 'splot' to de-convolve the \nii\ spectra, presenting \vrad, FWHM, and \vrad\ of the blue component.
  The lowest panels show flux calibrated surface brightnesses as derived from HST images and the ratio of surface brightness of the blue 
  component to that arising from the red component. In left to right order these data are discussed in Section~\ref{sec:NE}, Section~\ref{sec:BrightBar},
  and Section~\ref{sec:narrow} respectively.   
  }
     \label{fig:three}
\end{figure*}

 We present our results in the middle three panels of Figure~\ref{fig:three}. The vertical dashed reference line shows the position of the peak surface brightness in \nii\ S(\nii ). The top panel shows the variation in \Te\ (from \nii ) and \Ne\ (from \sii ) as derived from our treatment
 of the MUSE data. In the middle panel we show most of our results from the \nii\ spectra and the lowest panel shows the profile of the surface brightness
 in \nii\ and \oiii.  
 
 In the top panel we see that \Te\ begins at 8800$\pm$100 K on the northwest side and peaks at 9100$\pm$50 K at 2\arcsec\ beyond the 
S(\nii ) peak, then drops to 8700$\pm$100 at the end of the samples. 
In contrast, the \Ne\ begins at 2000$\pm$50 \cmq\ and rises to 3400$\pm$100 \cmq\ about 1\farcs5\ inside the S(\nii ) line,       
then, descends to 1600$\pm$50 \cmq. 

Since the \nii\ emitting layer is much thinner than the \oiii\ layer, we
only present the results from using 'splot" to de-convolve the 
\nii\ 658.3 nm line, the results of this appearing in the middle panel. 
Within the paradigm that variations in radial velocity of the MIF are due to variations in the tilt we can interpret the slow increase
in \vrad (\nii ) as an increase in the tilt passing a peak at 2\arcsec\ beyond the S(\nii ) peak, then dropping abruptly, and assuming
a constant lower tilt 3\arcsec\ beyond the S(\nii ) peak. 
On the inside of the \bb\ \vrad (\nii ) is 22.5 \kms, rising to a maximum at the \bb\ peak of 23.6 \kms, then dropping to 18.2 \kms\ 
outside the \bb. 

\subsubsection{3-D structure determined from \vrad (\nii)}
\label{sec:bb3D}

If the \nii\ expansion velocity 
is 10 \kms (Section~\ref{sec:Vcorr} and if the average \vrad\ for \cii\ and CO is 27.4 \kms\ (Section~\ref{sec:Vcorr}) is representative 
of the velocity of the underlying molecular cloud, then the
inner region is tilted 57\degr, the \bb\ 71\degr\  and the outer region 26\degr\ with respect to the plane of the sky. 
This means that the inner region is already quite tilted and the \bb~is only a locally greater tilt. The outer region is the least tilted.
These values are very sensitive to the adopted expansion velocity, which is less certain than the molecular cloud velocity.

If one assumes that the outer region is in the plane of the sky, then \Vexp (\nii) would 9.2 \kms\ rather than 10 \kms and
the tilt inside the \bb\ would be 53\degr\ and the \bb\ 68\degr\ with respect to the plane of the sky.  If the \nii\ expansion velocity is the minimum value of 9.2 \kms , the angles would be 33\degr, 24\degr, and 90\degr\ respectively. 
If the \nii\ expansion velocity is greater than 10 \kms\ then the angles would all be smaller. 

The inner region being highly tilted agrees with the 3-D model for the inner \hr\ constructed by \citet{wen95} from measurements
of the radio continuum surface brightness and densities derived from the \sii\ red doublet. The tilt at the \bb's peak is less than that
given in studies of comparing model ionized slabs with the observations and these differences are probably due to the expansion velocity used and a break-down of the assumed model in a region of rapidly changing tilt. The tilt of the MIF outside the \bb must not be as large
as 90\degr\ because the outer parts of the nebula would be shielded from the ionizing radiation from \tc. In fact, the shape of the inner \hr\ must
be determined by the photo-evaporation of material through the MIF as it eats its way into the host molecular cloud. 

\subsubsection{What the sequence of characteristics tells us about the 3-D structure of the \bb}
\label{sec:bbsequence}

We have a small but clear sequence of properties across our \bb\ sample.
The peak in \Ne\ occurs at 12\farcs0 from the \tc\ end of the sample, the S(\nii ) peak occurs at 12\farcs8,
the maximum \Te\ occurs at 15\arcsec , and 
the change of velocity characteristics \vrad (\nii) and FWHM occurs at 17\arcsec.
It should be recalled that all of these characteristics are derived from \nii\ observations 
except for \Ne, which is derived from observations of \sii.

The most straightforward interpretation is that the \Ne\ peak occurs within an \nzone\ layer viewed almost edge-on 
along a line-of-sight where there is the most favorable combination of events producing \sii\ emission (i.e.
the ionic abundance of S~$^{+}$, \Ne, and \Te).  The S(\nii) peak would be along a line-of-sight where most 
favorable combination of density of \Ne, N~$^{+}$, and \Te\ occur. The inner part of this \nzone\  layer would have
the highest total gas density as material is photo-evaporated away from the ionization boundary of the MIF. 
The \Te\ maximum would occur in the level of the \nzone\ layer where conditions for emitting \nii\ are best and also
where the local \Te\ is rising because of radiation hardening that occurs as one approaches the MIF and the increase 
of the gas density. Finally, the local peak in \vrad(\nii) and FWHM marks the point of inflection, where the \nzone\ layer
tilt changes abruptly. However, the expected relation between \vrad\ and FWHM (FWHM is larger where \vrad\ is smaller) is not seen.

\subsubsection{The Blue Layer emission near the \bb}
\label{sec:BBblue}

The Blue Layer \vrad\ changes abruptly at about 2\arcsec\ inside the position of the drop in \vrad (\nii). This argues that the Blue Layer 
is different on the two sides of the \bb. The Blue Layer component surface brightness mimics that of S(\nii) as one sees in the lowest panel,
where S(blue)/S(mif) does not change significantly when crossing the S(\nii) maximum. However the maximum Blue Layer surface brightness
[S(\nii)$\times$S(blue)/S(mif)] peaks at 13\farcs5,  near where \vrad (blue) peaks before dropping further out.  

This article cannot give a definitive answer about the origin, location, and nature of the Blue Layer even though it adds to the body of knowledge
about this feature. A more comprehensive paper is in preparation by co-author O'Dell.

\begin{figure}
 \includegraphics
 [width=\columnwidth]
 {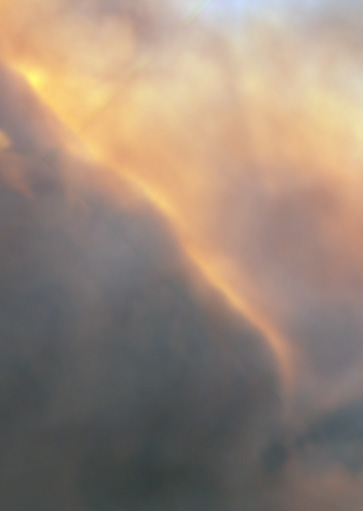}
    \caption{ This 14\farcs5$\times$20\farcs4 field is designated as the SW Bright Bar in Figure~\ref{fig:HSTfov}. 
    The colour coding is the same as in that figure, although the colour balance is slightly different.
    A 4\farcs0 sample along the linear portion in the southwest portion of this image was used in the study presented in the right-hand
    panels of Figure~\ref{fig:three}.  }
     \label{fig:SecBB}
\end{figure}

\subsection{A SW \bb\ Narrow Feature}
\label{sec:narrow}

In Figure~\ref{fig:HSTfov} we designate as the SW Bright Bar that section of the \bb\ that is distinctly different.
In this smaller region there is a wavy narrow feature that looks like a sharp ionization front that we designate as the Secondary Bright Bar. We have sampled a linear portion of this
feature by a series of 0\farcs5 samples 4\farcs0 wide with PA = 54\degr, with the resulting profiles shown in the right hand panels of Figure~\ref{fig:three}. The centre of the sharp feature, seen best in Figure~\ref{fig:SecBB} is at 5:35:19.64 -5:25:12.1. 

One has to bear in mind in examining the profile of \vrad, FWHM, and \vrad (blue) that these were made from spectra with a spatial resolution of about 2\arcsec\ whereas
the other data have better resolution (\Ne\ and \Te\ about 1\arcsec\ and the surface brightness data about 0\farcs1). 

The distribution of S(\nii) is similar to that of the \bb , but the peak is much sharper, which is quite consistent with Figure~\ref{fig:HSTfov} that 
shows the feature to be the narrowest structure along the \bb. The S(\oiii) variation is similar to that of the \bb\ in that it is more diffuse and 
peaks closer to the end of the sample facing \tc. The S(\nii) peak argues that we are seeing an ionization front almost edge-on.
Unlike the \bb\ both \Te\ and \Ne\ peak beyond the S(\nii) peak at 3\farcs9. The density peak of 4700 \cmq\ rises above the background of 2700 \cmq . The \Te\ peak of 9800$\pm$200 K lies above the closer region value of 9400$\pm$200 K and the further region value of 9000$\pm$200 K.

However, the \vrad(\nii) distribution disagrees with the model presented in the previous paragraph. The peak value (27.1 \kms at 1\farcs2) is well inside the S(\nii) peak and \vrad\ decreases along
the sample.  The peak velocity is almost the same as our adopted PDR velocity of 27.4 \kms\ and this indicates that the inner sample of the MIF is viewed almost edge-on and further from this peak the MIF is flattening. The most likely interpretation of the contradiction of the surface brightness, density, and temperature  with the radial velocity data is that the dynamic model (\vrad\ is determined by the tilt of the MIF) has broken down at 
this small scale and location. We also see an indication of this breakdown in that the inner FWHM  is comparable to the inner FWHM in the
\bb\ samples, even though \vrad\ is much higher.  Although the \vrad\ and FWHM values vary inversely with one another, in the outermost samples  the FWHM has become larger. This is again unlike the \bb. 

In contrast with almost all the other regions sampled, any backscattered red component is too weak to be detected. 
Although this is expected when one looks at an edge-on ionization front, in this case the absence is on both
sides of the S(\nii) peak. Another thing to be noted is that the large displacement \vrad(blue) values are well-defined
and measured accurately by ''splot'', but in the lower displacement samples the ''splot'' values plotted in Figure~\ref{fig:three}
are too large, as visual examination of the spectra indicates that those velocities should be about -3 \kms.

The Blue Layer emission velocity is the highest among our samples then drops  down to nearly the lowest velocity in our samples.
There is a general decrease (left to right) in the \nii S(blue)/S(mif) components, but there is a local rise immediately outside the S(\nii) peak.   
In summary, the Blue Layer emission behaves very much like the \bb\ in terms of a change of \vrad\ and S(blue)/S(mif).  

\subsection{The Northeast Sample}
\label{sec:NE}
\begin{figure*}
	\includegraphics
	[width=5in]
	{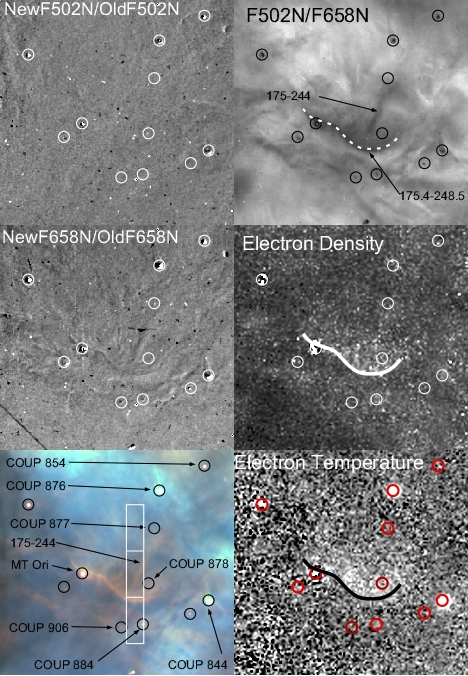}
    \caption{
 The 26\farcs9$\times$25\farcs9 Northeast Sample region (Figure~\ref{fig:HSTfov}) that includes the Northeast Arc (traced 
 with curved lines in the right hand panels is shown in multiple depictions as labeled. The colour coding in the bottom
 left panel is the same as in Figure~\ref{fig:HSTfov}. The other panels are explained and discussed in Section~
 \ref{sec:NE}. The open circles are proplyds, that are characterized by high \Ne, high \Te, and low F502N/F658N. The box in the bottom left panel shows the location of the series of 2\farcs0 samples at 0\farcs5 intervals taken and detailed in Figure~\ref{fig:three}.
  }
     \label{fig:NE}
\end{figure*}
We note in the maps of higher \Te\  that to the northeast of the Trapezium is a well defined arc, which we designate 
here as the Northeast Arc (NE-Arc). 
This feature falls within the Northeast Sample within Figure~\ref{fig:HSTfov} and is shown
in detail in Figure~\ref{fig:NE}. Although distinctive, it has not attracted previous attention. The southernmost position of the NE-Arc
is at 5:35:17.54 -5:22:48.5, which leads to its position based designation \citep{ode15}  as 175.4-248.5.

In Figure~\ref{fig:NE} we show the region around the NE-Arc. The lower left and upper right panels show that the feature is low ionization (\nii~is strongest)
and that it encloses a 5\farcs9 linear low ionization feature designated as 175-244. 
We note that the enhanced \Ne\ and \Te\ regions fall within the concavity of the NE-Arc. 
We searched for high proper motion objects within this FOV by aligning F502N and F658N images from the HST WFPC2 early observations (program GO 5469) and the more recent program GO 10921 observations, then taking their ratios. The results are shown 
in the upper two left-hand panels of Figure~\ref{fig:NE}. An object of high proper motion will appear as a double image, with a leading
bright and trailing dark structure. There are hints of motion, but no large tangential motions in the NE-Arc are obvious. 

\subsubsection{Characteristics of a north-south series of spectra crossing the NE-Arc}
\label{sec:profileNEarc}

In order to more quantitatively evaluate this feature, we sampled a 2\farcs0$\times$16\farcs0 region indicated in the lowest left panel of 
Figure~\ref{fig:NE} in a series of  2\farcs0 wide samples at spacings of 0\farcs5. The results from these spectra and WFC3 images of the same
region are shown in the left-hand panels of Figure~\ref{fig:three}. 

In Figure~\ref{fig:three} --left we see in the bottom panel that the NE-Arc resembles an ionization front viewed nearly edge-on for there is 
a sharp peak on S (\nii) at 5\farcs1 with a broader \oiii\ peak at 4\farcs5.  The region inside and outside the NE-Arc is highly structured, with much
more variation of surface brightness in \oiii. The top panel shows that \Ne\ peaks at about 6500$\pm$300 \cmq, well above a background of 3000$\pm$500 \cmq~and that \Te\ reaches a peak of about 10900$\pm$300 K, well above the southern side background of 9800$\pm$500 K and the northern background of 9600$\pm$200 K. The peak \Te\ occurs about 1\arcsec\ outside of the \Ne\ peak.  

The properties of the MIF lines vary significantly along the slit. South of the NE-Arc \vrad(\nii) is almost constant at 23.5 \kms,
then begins to drop at about the position of the S(\nii) peak to a low of 16.5 \kms, then recovers to a local peak of 20 \kms\ 
before continuing to drop further north. Within the model of the radial velocity being determined by the local inclination, the southern region is steep (about like the region north of the \bb, then flattens beyond the S(\nii) peak before becoming steeper and finally slowly flattening out. The pattern for the change in the FWHM is similar to that of the \bb, with there being a local peak outside of the S(\nii) peak, followed by a gradual decrease outward. 

An even bigger change across the S(\nii) peak is in the Blue Layer component. Its \vrad\ changes by
-15 \kms\ and its surface brightness as compared with that of the MIF's \nii\ emission drops by an order of magnitude. 

What is the NE-Arc feature, a shock like HH~204 or a structure within the MIF of the nebula, like the \bb?
The \Te\ and \Ne\ variations more closely resemble the \bb, in that the HH 204 peaks occur ahead of the S(\nii) peak. 
There is no evidence for a high velocity flow, either tangentially or along the line-of-sight, thus it is not like HH 204. 
Finally, the variations is the velocity and FWHM change rapidly behind the S(\nii) peak, like the \bb\ and unlike HH 204. 
Finally, there is evidence for significant changes in the Blue Layer velocity across the S(\nii) feature. All of thisIn the lower-left panel
argues that the NE-Arc is a structure within the MIF and represents a region were the photoionized front has encountered
an underlying spheroidal high density concentration in the underlying molecular cloud.

\subsubsection{The origin and consequences of the 175-244 feature}
\label{sec:175}

We have looked for an association of the narrow feature 175-244 with the star COUP 854 (a young star of spectral type K0-K2,  \citep{hil97}), and the NE-Arc. 
The HST images show that the star has a faint low ionization structure 0\farcs3 at PA~=~98\degr\ that appears to curve south
to 0\farcs4 at PA~=~131\degr. The position of the NE-Arc relative to COUP 854 is 17\farcs2 at 152\degr. Feature 175-244
is oriented to PA~=~155\degr\ in its southernmost section,  but curves towards COUP 854 in it's northern portion. The NewF658N/OldF658N image in Figure~\ref{fig:NE} indicates that the more southern portion of 175-244 is moving to the 
westsouthwest, but at little more than the uncertainty of a few kilometers per second.  A search of the \nii\ spectra in this area reveal no red or blue components that cannot be more simply assigned to the backscattered red components or the emission from the Blue Layer.

The \nii\ features very near COUP 854 do not resemble jets coming from that star and their orientation is quite different from the angle from the northernmost portion of 175-244 to the star (148\degr). It is very unlikely that feature 175-244 is part of a collimated outflow from the star.  Since the final direction of the 175-244 feature (155\degr) does not agree with the axis of the NE-Arc in this region 166\degr, it is unlikely that this is a jet driving the NE-Arc. 

However, there is some evidence for an interaction of the 175-244 feature with the MIF. As noted in the previous section, \vrad(\nii) drops to a local low of 16.5 \kms\  and that occurs at 7\farcs2. This is also the position that 175-244 crosses our series of sample spectra and at the position of the lowest S(\oiii) in our sample. In addition, \vrad(Blue) shows a local high at this point
as does the ratio S(blue)/S(mif). These characteristics indicate that the 175-244 feature penetrates into an important region of \oiii\ emission (probably near the \nzone\ to \ozone\ boundary) and affects the velocity of the \nzone\ layer.

\subsection{An Ionization Shadow}
\label{sec:shadow}
\begin{figure}
	\includegraphics[width=\columnwidth]{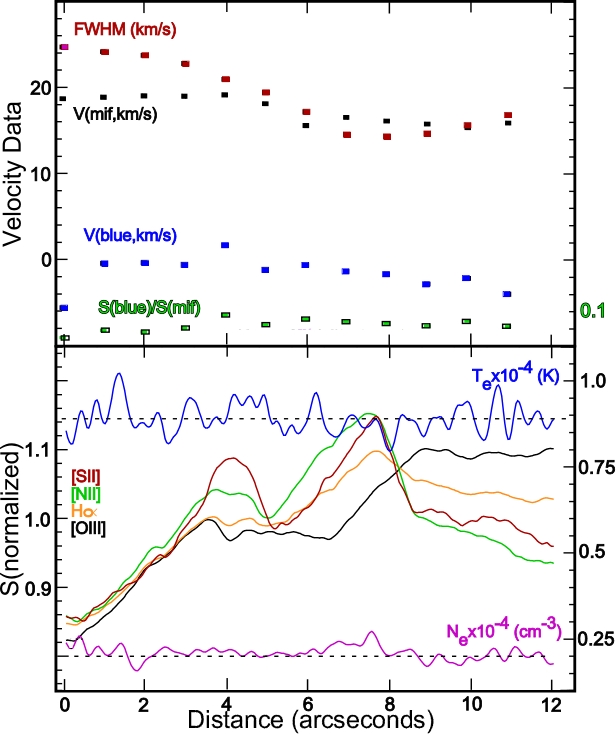}
    \caption{
    This figure shows the results of a profile across the dark shadow at PA~=~195\degr\ from \tc\ as shown in Figure~\ref{fig:HSTfov} and described in Section~\ref{sec:shadow}. Distance progresses towards PA~=~287\degr. In the upper panel the \nii\ \vrad\ and FWHM information is given using the coordinate values 
    on the left ordinate. The S(blue)/S(mif) values are scaled according to the coordinate values on the right ordinate.
    In the lower panel the results for the normalized emission-line surface brightnesses are shown, scaled to the coordinate values
    on the left. In the same lower panel the \Te\ and \Ne\ values are shown in units of 10$\rm ^{4}$ using the right-hand coordinate values.    
    }
     \label{fig:shadow}
\end{figure}

There are a number of radial features oriented towards \tc\ and models for them predict that their \Te\ values should
be lower than the surrounding gas. We have investigated one of the best defined of these features in order to assess 
their role in temperature variations across the \hr. 

These objects were first noted in both the \hr\ and NGC~7293 (the Helix Nebula) through images in S(\oiii)/S(\nii) where they 
appear as a pair of radial lines centred on the ionizing star and being dark in between \citep{ode00a}.
They are essentially shadow cones 
behind objects that are optically thick to the Lyman Continuum (LyC). The shadowing objects in the \hr\ are the proplyds.
The shadows are only evident if there is ambient material that is photoionized and they can be used to trace the 3-D structure
of the nebula \citep{ode09b}. Within the shadow cone the ambient gas is illuminated by LyC radiation formed by recombinations of surrounding
ionized hydrogen, rather than \tc. This radiation field is about an order of magnitude less than that from the star, but is
sufficient to cause an ionization front to form on the shadow of the cone. 
The illuminating LyC photons have a flux distribution more concentrated near the ionization threshold for hydrogen 
and the photoionized gas on the shadow cone's surface will have a lower electron temperature. This model created by \citet{can98} predicts that
a profile across a shadow cone will show a low ionization boundary (sharp through limb brightening) with the region between
these boundaries be marked by seeing the cone's ionization front from face on and missing high ionization emission from 
the ambient gas within the shadow cone.

This is the same process that determines the shape of the tadpole shaped tails of the proplyds (Bally, O'Dell, \&\ McCaughrean 2000). The important difference
is that in the case of the proplyds the density of gas in the shadow is determined by the rate of material being lost
through photoionization from the side facing \tc\  and this density decreases with increasing distance from the source.
This decreasing density allows the shadowed zone's ionization front to advance into the centre of the shadow and 
give the proplyds their characteristic form.

We selected the shadow cone whose centre is at PA~=~195\degr\ from \tc. Figure~\ref{fig:HSTfov} shows the position
of this feature and the sample box taken crossing it. The boundaries cross through the proplyd AC Ori most recently 
discussed by \citet{ode15}.  The sample box was 3\farcs5$\times$12\farcs0 and oriented towards PA~=~287\degr\ and the shadow is slightly more than 2\farcs0 wide. It was examined 
both in WFPC2 images in \oiii, \nii, \Ha, and \sii\ and also in Spectral Atlas \nii\ spectra in steps of 1\farcs0. 

The results of the profile are shown in Figure~\ref{fig:shadow}. The centre of the shadow is at 6\farcs2. We see that
S(\oiii) increases with distance and dips at the position of the shadow. This is explained as a lack of emission in \oiii\ 
in shadow cone passing through ambient doubly ionized oxygen.  The \nii\ background rises as one approaches the 
shadow from either direction, peaks just inside of the \oiii\ dip as one sees the limb brightened ionization boundary on the outside
of the shadow, then it dips down as one sees emission from the layer flat-on. The \sii\ profile peaks indistinguishably at
the same position as \nii, which reflects the fact that the shadow's ionization zone is about 1\arcsec\ thick. The \Ha\ surface
brightness illustrates how it arises from all the ionized gas. The radial velocity slowly decreases across the profile and there is
only a hint of a local increase at the edges of the shadow's ionization front, but the spatial resolution of the spectra precludes
any interpretation. The same can be said of the large change in the FWHM, although it appears to dip across the shadow. 
There is only a slow decrease in \vrad (Blue) and increase in S(blue)/S(mif) across the shadow, indicating that material affected
by the shadow is not part of the Blue Layer.

The backscattered component is not detectable in the first four samples (0\arcsec\ through 3\arcsec), but then begins
to be stronger relative to the MIF component, until becoming stronger than in most of our other samples. This absence 
occurs at positions where the FWHM is high. 

The derived \Ne\ increases from the local value of 2000$\pm$200 \cmq\ to about 2300$\pm$200 \cmq\ at the more distant limb brightened
\nii\ edge of the shadow. This is interpreted as that edge being of higher density than the ambient more highly ionized gas. 
There is only a hint of a dip in \Te\ across the shadow from the local value of 8900$\pm$300 K.

\begin{figure*}
	\includegraphics
	[width=7in]
	{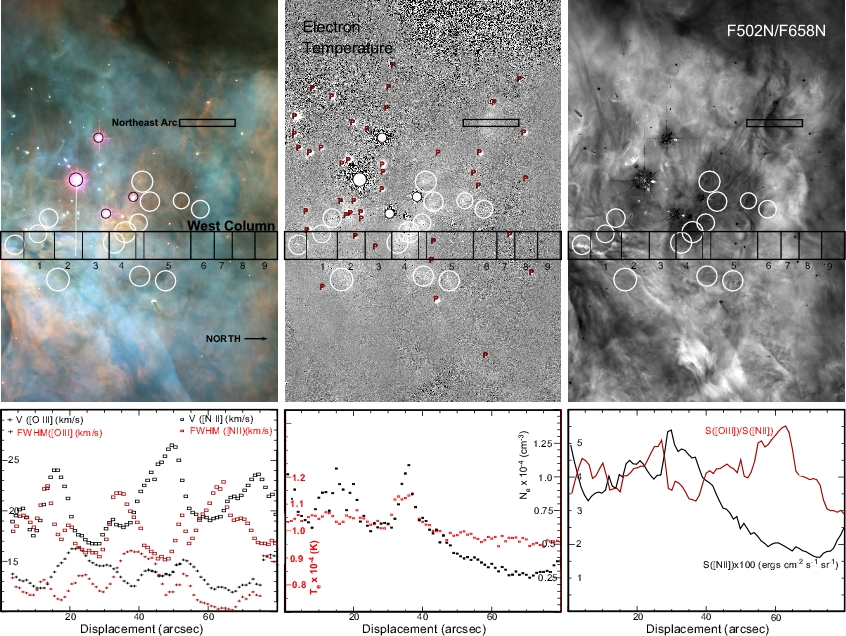}
    \caption{Each top panel shows a 115\farcs6$\times$79\farcs7 field of view, as shown in Figure~\ref{fig:HSTfov}, except now north is toward 
    the right and east is toward the top. 
    The left top panel is the WFPC2 image as colour coded in Figure~\ref{fig:HSTfov}. 
    The centre top panel is a map of \Te\ determined from \nii. 
    The right top panel shows the ratio of the images of \oiii\ (F502N) over \nii\  (F658N) uncalibrated WFPC2 images.
    For reference, the sample region for the Northeast Arc, discussed in Section~\ref{sec:NE} is shown in each panel.
    Each top panel also shows a column of south-north samples as described in Section~\ref{sec:NorthernArea}.
    The white circles indicate resolved regions of high \Te\ and the red `P' letters indicate unresolved high \Te\ regions associated with proplyds.
    The lower panels present results for the north-south column in the upper panels taken at 1\arcsec\ intervals as described in Section~\ref{sec:NorthernArea}.
    The numbered groups (1--9) within the West Column are those used in the discussion and in Table~\ref{tab:WestColumn}.
 }
     \label{fig:northern}
\end{figure*}
   
\begin{landscape}
\begin{table}
\centering
\caption{Properties of Samples within the West Column}
\label{tab:WestColumn}
\begin{tabular}{lccccccccc}
\hline
                              & Group 1                            &Group 2               & Group 3              &Group 4           & Group 5           & Group 6          & Group 7        & Group 8             & Group 9\\
 \hline
Range (arcsec)     &6--15*                                 & 15--23                  & 23--31              & 32-39               & 41--54*            & 55--61            & 62--66*          & 67--73*              & 74--80*\\
\vrad(\nii) (\kms)    & 20.7$\pm$2.6                   &19.3$\pm$2.5      &17.7$\pm$1.0      &19.9$\pm$1.7    &23.5$\pm$2.4  &19.3$\pm$0.1  &21.0$\pm$0.7  &22.8$\pm$0.8  &19.8$\pm$1.9\\
FWHM(\nii) (\kms) &19.5$\pm$1.8                    &17.0$\pm$1.0      &17.7$\pm$2.5      &19.6$\pm$1.8    &16.7$\pm$1.4  &20.8$\pm$1.3   &20.2$\pm$1.5    &17.8$\pm$0.9  &16.4$\pm$0.5\\
\vrad(\oiii) (\kms)   &14.0$\pm$1.0                    &16.5$\pm$0.6      &15.3$\pm$0.3      &13.9$\pm$0.5     &14.9$\pm$0.5  &13.4$\pm$0.5    &12.9$\pm$0.2      &13.9$\pm$0.9  &16.6$\pm$0.7\\
FWHM(\oiii) (\kms)&12.9$\pm$0.7                    &13.4$\pm$0.9      &13.6$\pm$0.8      &16.5$\pm$0.3      &13.6$\pm$1.2   &11.3$\pm$0.3    &11.2$\pm$0.8       &12.9$\pm$1.5  &16.5$\pm$1.3\\
\vrad(blue,bluer) (\kms)       &-13.9$\pm$2.7      & -6.2$\pm$3.0    &-9.6$\pm$1.7        &-9.8$\pm$5.6      &  -8.8$\pm$5.0    &-11.4$\pm$3.9   & -14.0     &  ----               &    ---- \\
\vrad(blue,redder) (\kms)     &   7.0$\pm$2.3      & 6.1      & ----                       & 3.1$\pm$0.8       &   9.7$\pm$1.0     &  ----                   &  1.2$\pm$01.5   &  4.9$\pm$3.2 &  0.7$\pm$2.9\\
S(blue,bluer,\nii)/S(mif,\nii)  &0.016$\pm$0.003 &0.029$\pm$0.006&0.019$\pm$0.010&0.013$\pm$0.005&0.014$\pm$0.003&0.021$\pm$0.006&0.015   &  ---                    & ----\\
S(blue,redder,\nii)/S(mif,\nii)&0.050$\pm$0.009&0.054      &  ----                     & 0.058$\pm$0.024&0.117$\pm$0.036&       ----              &0.049$\pm$0.006&0.076$\pm$0.025&0.044$\pm$0.013 \\
\Ne (\cmq)             &8450$\pm$1250&8360$\pm$1050&6460$\pm$390 &8960$\pm$2000&4930$\pm$930&3610$\pm$250  &3020$\pm$220   &2820$\pm$120  &3240$\pm$300\\
\Te (K)                   &10520$\pm$100&10520$\pm$140&10270$\pm$160&11090$\pm$280&9960$\pm$160&9740$\pm$40    &9740$\pm$180   &9560$\pm$160  &9540$\pm$80\\
S(\nii)$\times$100 (ergs 
cm$\rm ^{-2}$ \pers 
ster$\rm ^{-1}$)     &3.62$\pm$0.25     &4.29$\pm$0.18&4.50$\pm$0.59 &  4.54$\pm$0.29&2.92$\pm$0.50&2.03$\pm$0.10&1.88$\pm$0.07&1.70$\pm$0.10&1.99$\pm$0.29\\
S(\oiii)/S(\nii)         &3.95$\pm$0.32     &4.15$\pm$0.27&4.41$\pm$0.56 &  3.54$\pm$0.22&4.22$\pm$0.16&5.08$\pm$0.16&5.05$\pm$0.56&3.93$\pm$0.09&3.03$\pm$0.13\\
\hline
\end{tabular}\\
*No backscattered red \nii~component in slits 11\arcsec --14\arcsec , 41\arcsec --50\arcsec , 63\arcsec --76\arcsec .
\end{table}
\end{landscape}

\subsection{Northern Area}
\label{sec:NorthernArea}

The structure of the region near the Trapezium stars is discussed in detail in an earlier publication \citep{ode09b}. There it is argued that 
the appearance of this region is determined not simply by processes occurring near the MIF, but also by a high-ionization shell of gas 
produced by the stellar wind arising from \tc.  This shell appears as an incomplete high ionization arc (HIA), being open to the south-west where the Orion-South cloud is located. The inner boundary of the shell corresponds to the boundary of the unimpeded stellar wind with the shocked outer region of that wind. The outer boundary would represent where the outward moving shocked wind gas interacts with the low density ambient gas. \citet{ode09b} argue that because there is not a surrounding \nii\ boundary (best examined in their wider FOV Figure 3) this shell is not ionization bounded.

The image in Figure~\ref{fig:northern} includes the north-west through east-southeast portions of the HIA. It appears
most clearly as a bright and broad feature in the F502N/F658N (top-right) panel of Figure~\ref{fig:northern}. The low F502N/F658N values 
in the innermost region and outside of the HIA are interpreted to be due to the dominance of MIF emission. In contrast, the 
low values to the south-west are due to seeing the north-east boundary of the Orion-South cloud in projection, much like the
\bb\ \citep{ode09b,mesa11}.

We have prepared a series of eighty samples each 8\farcs0 wide and 1\farcs0 high along a south to north line centred 19\farcs0 west of \tc, with
the southern limit 21\farcs6 south of \tc, as shown in each of the top panels of Figure~\ref{fig:northern}. The lower panels present 
radial velocity and FWHM for the strong \nii\ and \oii\ components determined using `splot' (left panel), our \Ne\ and \Te\ values (centre panel),
and the calibrated S(\nii) and the S(\oiii)/S(\nii) ratios from calibrated WFPC2 images (right panel). This series of samples begins at the northeast corner 
of Orion-South, proceeds through the putative centre of the effects of \tc 's stellar wind, then crosses the northern boundary of the high ionization incomplete arc. In all of the upper panels the location of known proplyds have been marked by a red letter P and circles
indicated the position of resolved high temperature regions not associated with a known proplyd.

In the lower left panel of Figure~\ref{fig:northern} we see that the the \nii\ velocities are more positive than the \oiii\ velocities, which is 
consistent with the basic model of this region being a photo-evaporating blister with the \oiii\ emitting layer having
a larger evaporation velocity than the \nii\ emitting layer (Section~\ref{sec:Vcorr}). 
Beyond this general observation, one must examine all the characteristics at the same place in the West Column. For this reason we have grouped the samples according to their characteristics into a sequence of larger groups and discuss the properties of these groups in the following sections.  The numeric results for each group are given in Table~\ref{tab:WestColumn}.

There is a loose correlation (coefficient 0.92) of \Ne\ and S(\nii) following the relation \Ne = -450+1.96$\times \rm 10^{5} \times$ S(\nii).
The former relation indicates the result of the areas of highest surface brightness being those that also 
experience the highest rate of photo-evaporation of material. Since the S(\nii) values are not corrected for extinction, the multiplier
in this equation would be lower when using extinction-corrected values of S(\nii).

We see numerous local variations of \Te\ and explain the cause of the largest in Section~\ref{sec:NeTe}. 
We see temperature fluctuations of about 400 K across the West Column with size scales of about 5\arcsec, with size
scales increasing and the magnitude of the temperature fluctuations decreasing as one moves away from the
point of the West Column that is closest to \tc.

The \vrad(blue,\nii) values within an individual group often shows a grouping into bluer (about -10.5$\pm$2.8 \kms) or redder (about 4.7$\pm
$3.2 \kms),
with the redder component being stronger [S(blue)/S(mif)=0.064$\pm$0.026] than the blue [S(blue)/S(mif)=0.018$\pm$0.006].

\subsubsection{The High Ionization Arc Region}
\label{sec:HIA}
Our Groups 6--8 cross the HIA. We see a local enhancement of the S(\oiii)/S(\nii) ratio and the \vrad(\oiii) values are lower
than the adjacent groups, indicating that the MIF \oiii\ emitting zone in these groups is flatter or that there is a negative 
velocity component of the partial shell that forms the HIA. Given the enhancement of \oiii\ emission associated with
the HIA, the latter interpretation is more likely to be correct. The jump in \vrad(\oiii) and FWHM(\oiii) in Group 9 strengthens
this interpretation. The abrupt change in \vrad(\nii) from Group 5 (23.4$\pm$2.0 \kms\ to Group 6 (19.7$\pm$0.6 \kms) and Group 7 
(20.0$\pm$1.0 \kms) argues that the MIF \nii\ emitting layer has been flattened behind the HIA and has recovered its
original orientation by Group 8 (22.4$\pm$0.6 \kms)  before flattening slightly more by Group 9 (19.7$\pm$1.7 \kms). 
It may be important that no backscattering component can be seen in samples 63\arcsec --76\arcsec, on the outer boundary of the HIA.

The northern-most samples (Group 9) are expected to be unaffected by the HIA and behave according to the \vrad model in Section~\ref{sec:Vcorr} if it is indeed a shocked shell of gas.
The low ratio of S(\oiii)/S(\nii) (the lowest in our nine groups) certainly indicates this to be the case. 
However, Figure~\ref{fig:northern}'s lower left panel shows that FWHM(\nii) is almost constant while \vrad(\nii) is dropping
rapidly. These values for \oiii\ are almost equal and show no change across the sample. 

\subsubsection{The Inner \nii\ Region}
\label{sec:InnerNII}

The Inner \nii\ Region is a distinctive feature of low ionization found surprisingly close to \tc\ and is most visible in the
F502N/F658N image in the top right panel of Figure~\ref{fig:northern} and it covers 32\arcsec\ -- 54\arcsec, with Group 4  well within the region of higher S(\nii) falling within the HIA and Group 5 being transitional from it to the 
characteristics of the HIA. It is probably the highest density (mean = 8960$\pm$2000 \cmq, peak = 11300 \cmq) and is the highest temperature (mean = 11090$\pm$280 K, peak = 11200 K) of our groups.
 
In Group 4  \vrad(nii) is low (19.9$\pm$1.7 \kms) and the FWHM(\nii) is high (19.6$\pm$1.8 \kms) and these reverse in 
Group 5 (\vrad(\nii) = 23.5$\pm$2.4 \kms, FWHM(\nii) = 16.7$\pm$1.4 \kms). This indicates that Group 4 is flatter than Group 5
The \vrad(\oiii) and FWHM(\oiii) behave in the same manner (Table~\ref{tab:WestColumn}. However, 
S(\nii) is high in Group 4 (where S(\oiii)/S(\nii) is low) and then in Group 5 S(\nii) has dropped while S(\oiii)/S(\nii) has increased.
The \vrad\ and FWHM properties agree with the pattern described in Section~\ref{sec:Vcorr}. The pattern of the surface brightness 
does not. Within Group 5 most of the samples (41\arcsec --50\arcsec) have no detectable backscattered component. 

It is likely that the low ionization of Group 4 is caused by the lower ionization factor U, the ratio of the flux of ionizing photons to the density of neutral hydrogen \citep{agn3}. This is because the flux of 
ionizing photons is essentially the same in both regions (only somewhat lower in Group 5), but the density is much higher in Group 4.  

\subsubsection{Properties of the Southernmost Samples of the West Column}
\label{sec:southern}

Groups 1--3 sample the transition region between the Inner \nii\ Region and the northeast boundary of the Orion-South.
When examining their statistics  (Table~\ref{tab:WestColumn}) they show little variation, except for the low \Ne\ for Group 3.
However, the devil is in the detail as one sees wide fine-scale fluctuations in most of the observed and derived quantities.

Using \vrad(\nii) as a guide, we can trace the changes of tilt across these three groups. The MIF must almost be flat at 
25\arcsec, then monotonically increases in tilt to a high angle at 14\arcsec, then decreases again to the south. There is no detectable
backscattering component in samples 11\arcsec --14\arcsec, which is in the area of decreasing angle of the MIF. The curve of 
the rise and decay from the 14\arcsec\ peak is mimicked by a much noisier rise and decrease in \Ne.  We are probably seeing
the effects of a varying flux of ionizing photons in a rapidly rising, then decreasing escarpment in the MIF. This pattern is rendered clearer when considering the 
range covered within the \vrad(\nii) peaks at 14\arcsec\ to 47\arcsec. The \Ne\ rises occur on the inside side of the two \vrad\ peaks. 
This argues that there we are seeing a concave pocket in the MIF with a bottom at 25\arcsec\ and the greatest angles at 14\arcsec\ 
and 47\arcsec. The continuation of high \Ne\ south of the 14\arcsec\ velocity peak and not to the north of the 47\arcsec\ peak 
indicates that the 14\arcsec\ peak MIF is physically closer to \tc\ and not just closer in projection on the sky.

\section{The Outflows Area}
\label{sec:outflows}

\begin{landscape}
\begin{figure}
	\includegraphics
    [width=9in]
    {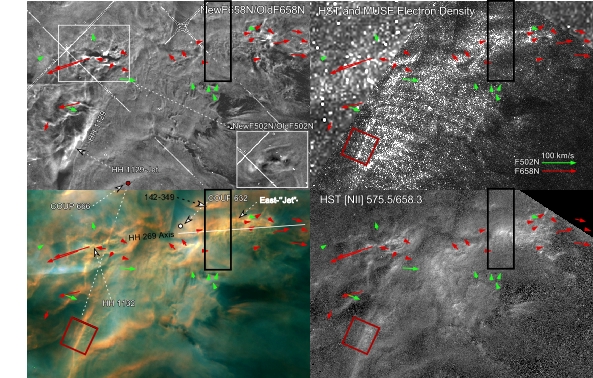}
    \caption{The 32\farcs6$\times$21\farcs7 panels in this array are all centred on 5:35:12.47 -5:23:58 (2000) as shown in Figure~\ref{fig:HSTfov}. Each panel is described in detail in Section~\ref{sec:outflows}. The black rectangle and red square indicate 
    the position of samples discussed in Section~\ref{sec:north} and Section~\ref{sec:SE}.}
     \label{fig:outflows}
\end{figure}
\end{landscape}

The region shown in Figure~\ref{fig:outflows} contains the sources of many of the large and small-scale stellar outflows in the \hr\ and has most recently been discussed in detail by \citet{ode15}. The lower-left panel uses the same colour coding as Figure~\ref{fig:HSTfov}.
The arrows in all of the images indicate the magnitude and direction of tangential motions (red \nii\ , green \oiii) using images from HST programs GO 5469 
and GO 10921, as described in Section~\ref{sec:NE}. The young star COUP 632 is almost certainly the source of the rapidly moving irregular
feature HH~1132. COUP 666 lies just north of the FOV and has been drawn as a red filled circle on the upper-left panel, although
its coordinate system is that of the lower-left panel. This star is likely to be the source of the narrow feature labeled as the HH~1129-Jet. This feature 
lies on a projection of a straight line from COUP 666
 through a narrow linear feature crossing HH~1132, but passes east of a nearly
parallel linear feature labeled as HH~1129. The white line in the lower-left panel show the nominal axis of a series of shocks
designated as HH~269. The source of the HH~269 shocks must lie near this line (or its
projection eastward) as discussed in \citet{ode15}.  The upper left panel illustrates both the motions of objects and also the change
in surface brightness of some figures (especially around HH~1129). 

The upper-right panel gives the results of deriving \Ne\ using the red \sii\ doublet line ratios. The upper left portion of this panel is from the MUSE
data set, while the remainder is from HST WFC3 program GO 12543. That program imaged the southwest portion of this FOV with
the narrow-band quadrant filters for the \sii\ lines. The quadrant filters are each of slightly less than one-fourth the area of the
standard WFC3 filters, but cover more diagnostic lines. The calibration of these filters are described in detail in a study in process by 
W. J. Henney.  The low resolution MUSE \Ne\ maps were carefully merged with the 10 times better resolution WFC3 data. 
The series of arcs pointing northwest in the lower part of the \Ne\ map are caused by uncorrected filter-fringing and are not important for 
our discussion.

The lower-right panel shows the ratio of calibrated \nii\ 575.5 nm and \nii\ 658.3 nm emission-lines from the full area GO 12543 WFC3 images.
This image is shown (rather than a \Te\ map because it provides the highest angular resolution and best S/N ratio. 

Figure~\ref{fig:outflows} allows one to relate the apparent features to their motions, changes, and conditions. We present here
a detailed analysis only for two regions of particular interest. The North Sample is shown as a black rectangle in Figure~\ref{fig:outflows}
and is discussed in Section~\ref{sec:north}. The Southeast Sample is shown as a red square in Figure~\ref{fig:outflows} and 
is discussed in Section~\ref{sec:SE}.

\subsection{The North Sample West of COUP 632}
\label{sec:north}
\begin{figure*}
	\includegraphics
	[width=6in]
	{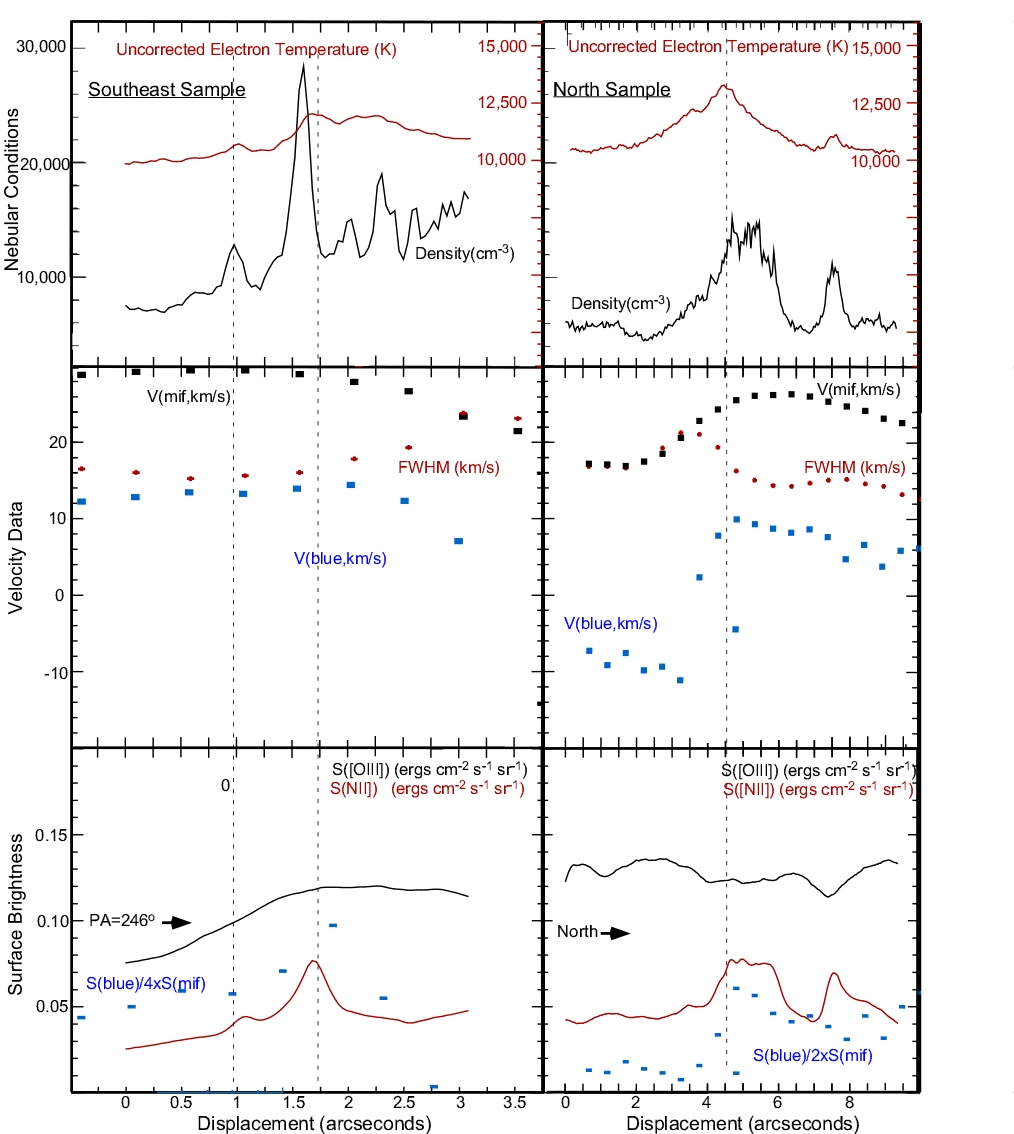}
    \caption{Both the left and right panels of this figure show sets of data as in several preceding figures,
    but are for the North (black box) and Southeast (red box)  samples drawn in Figure~\ref{fig:outflows}. 
    These are explained and discussed in Section~\ref{sec:north} and Section~\ref{sec:SE}. The vertical dashed lines are 
    for reference.}
     \label{fig:geheimnis}
\end{figure*}

This region was selected for study because it crosses the axis of the HH~269 flow and many irregular east-west features.
The sequence of irregular features about 2\arcsec\ wide at this point have been designated \citep{ode15} as the West-''Jet''. The quotation marks are added here because it is not established that the features represent a collimated flow. The lower part of the feature lies along the nominal axis of the HH~269 flow. The uncertainty of the PA of the axis of the HH~269 flow
is about 1\degr, which corresponds to about $\pm$1\arcsec\ in declination. The West-''Jet'' does not reveal high tangential velocities until its western end.
This region is also one of the hottest areas of the \hr.
The FOV is 3\farcs0$\times$9\farcs3 and the \nii\ spectra were taken in samples of 3\farcs0 wide by one Spectral Atlas line (0\farcs53) 
high. The calibrated surface brightnesses for \nii\ and \oiii\ from the HST images are shown in the right-hand panels of Figure~\ref{fig:geheimnis}. 

We have also presented 
the results of the \sii\ density and uncorrected-\Te\ profiles in the upper-right panel. 
Both of these require explanation. The critical density (the density at which it is equally likely that the upper state is depopulated by spontaneous radiative decay or a collisional de-excitation) for the \sii\ doublet lines are about 1000 \cmq. The critical densities are different
for the two upper levels giving rise to the doublet, which make it the useful \Ne\ indicator that it is. However, the low value of
the critical densities means that a derived \Ne\ becomes much more sensitive to photometric errors with increasing density and derived densities of more than about 10000 \cmq\ are very uncertain \citep{agn3}. We see this in the bright points in Figure~\ref{fig:outflows} and noisy peaks of \sii\ derived values of \Ne\ shown in Figure~\ref{fig:geheimnis}.  

The \Te\ values in this figure are labeled as ''Uncorrected''. This is because they have not been corrected for collisional de-excitation 
of the upper electron state giving rise to the \nii\ 658.3 nm line. This de-excitation supresses the 658.3 nm line emission and makes the 
derived \Te\ value too large. Regions with \Te\ about 11000 K and density about 5000 \cmq\ will be over-estimated about 3\%. The over-estimation increases
to 6\%\ at 10000 \cmq\ and 8\%\ at 15000 \cmq. This means that the peak values of \Te\ are probably too large by about 1000 K and that
the narrow peak in \Te\ at displacement 7\farcs5 can be explained as primarily due to collisional de-excitation.  However, 
this analysis leaves intact that there is a very high \Te\ (about 12000 K) at displacement 4\farcs5.


We see in Figure~\ref{fig:geheimnis} that \Ne\ and \Te\ begin and end at about the same values, 3500 \cmq\ and 10450$\pm$150 K, respectively. 
However, significant increases arise in both, beginning at about 2\arcsec. A local peak in \Te\ occurs at 4\farcs5, then it gradually decays
to the north, with a local rise at 7\farcs5 that is explained above as being due to a local increase in \Ne. The electron density essentially peaks at the \Te\ peak
(4\farcs5), but is about constant until 6\farcs0, before dropping to the background value, interrupted by a strong increase at 7\farcs5. 
The central region (4\farcs5 to  6\farcs0) of both high \Ne\ and \Te\ occurs where the samples transit the West-''Jet''. The 7\farcs5 \Ne\ peak occurs at 
an east-west oriented low ionization feature (5:35:14.21 -5:23:48.9, 142-349 in shorthand) that curves northward on its east end.

S(\nii) rises to a flat peak that agrees in position with the passage of the West-''Jet'', then drops to the background level before peaking at 
7\farcs5, then returning to the background level. Across the full sample S(\oiii) varies little, except for a small dip from 4\farcs0 to 6\farcs0, and a sharp and stronger dip at 7\farcs3.  

This 
information argues that the West-''Jet'' lies within the \oiii\ emitting zone near the MIF of the Orion-South cloud that faces the observer. The West-''Jet''  is of low ionization, having no emission of its own in \oiii\ and excludes only a small amount of \oiii\ zone emission. 
The fact that the 142-349 feature is more than double the local density yet has a surface brightness only one-third higher indicates that this feature is a small fraction of the thickness of the \nii\ emitting layer. We also know that it must be located in the \oiii\ emitting zone since that emission
dips at the \nii\ bright position. It must be closer to the MIF than the West-''Jet'' since it excludes a larger fraction of the \oiii\ emission.

\vrad (\nii) is quite low (about 14 \kms) at the southern end of our sample, indicating that this region is nearly flat-on. This velocity increases to a broad peak of 22 \kms\ at 6\farcs0, then slowly decreases to 18 \kms. This velocity change across the 
position of the West-''Jet'' is certainly broader than that feature. However, when the Spectral Atlas resolution of 2\arcsec and the location of the 142-349 feature are considered, the \vrad (\nii) change can be attributed to either the MIF being more tilted there or that the West-''Jet'' has a more positive radial velocity. If the West-''Jet'' is the driving source of HH~269, then the velocity change is probably due to an increasing tilt of the MIF since
the HH~269 shocks are known to be blue-shifted \citep{ode92}.

The FWHM(\nii) is nearly constant at about 12 \kms, except for a sharp rise that peaks at 18 \kms\  (peak at 3\farcs5, with boundaries at about 2\farcs0 to 4\farcs5). Examination of Figure~\ref{fig:outflows} shows that this region of unusually large FWHM is where 
one sees multiple fine-scale tangential motions, indicating small scale motions within the MIF \nii\ emitting zone. The disturbed region runs from
3\farcs0 to 5\farcs0, indistinguishable from the FWHM anomaly. This region is on the south edge of the West-''Jet'' and is probably 
caused by it. 

There is a linear series of low tangential velocity knots east of displacement 4\farcs2. They extend from 1\farcs0 to 4\farcs0 from the east boundary of 
the North Sample and have PA=275$\pm$2\degr, indistinguishably the same as the axis of HH~269. This argues that the south border of the West-''Jet'', where \Te\ peaks, marks the collimated flow that drives HH~269.

Both \Vblue\ and S(blue)/S(mif) change abruptly from nearly constant values at the lowest displacements to about 17 \kms\ more positive
velocity and five times stronger surface brightness ratio at the same point where the \Te\ maximum occurs. 

The pattern of two preferred blue component velocities at near -10 \kms\ and +5 \kms\ has already been seen
in the West Column data (Section~\ref{sec:NE})  and the Northeast Arc (Section \ref{sec:NorthernArea}). Similar to the case with 
the North Sample, those other regions have the higher radial velocity blue component stronger than the lower velocity counterpart. The higher displacement values of \Vblue\ are somewhat higher than the characteristic value of 1.8$\pm$1.9 \kms. The abrupt
changes of the \Vblue\ component indicate the the Orion-South cloud influences the Blue Layer after the MIF has tilted towards the observer along the northern boundary of the West-''Jet'' feature. 

\subsection{A Newly Identified Feature in the HH~1149-CCW Structure}
\label{sec:SE}

A fan of shocks emanating immediately to the east of the Orion-South cloud and designated as HH~1149 was recently identified \citep{ode15}. On its most counter-clockwise (CCW) boundary is an irregular structure designated as HH~1129. At a point immediately north of the red box in Figure~\ref{fig:outflows} HH~1129 changes PA by a few degrees CCW and retains its broad structure. There is a newly discovered feature, which we call the
HH~1129-Jet, which parallels HH~1129 and lies exactly on a line projecting back along the northern parts of HH~1129, through the jet feature 149.1-350.2 that crosses HH~1132, and then reaches COUP 666 (Section~\ref{sec:outflows}). These two features are quite different,
with the HH~1129-Jet being straight, narrower, and of lower ionization.

\begin{figure}
\includegraphics[width=\columnwidth]{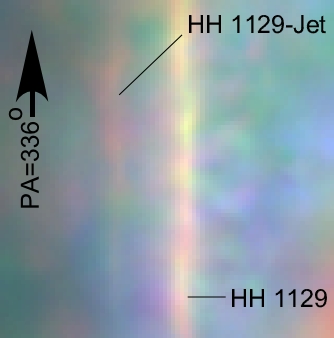}
 \caption{
This 3\farcs20$\times$3\farcs24 image is of the SE Sample outlined in red in Figure~\ref{fig:outflows}. 
The colour coding is the same as in the previous colour images, but the colour balance has been changed for clarity.}
\label{fig:SEcolor}
\end{figure}

We show a sample that includes both the HH~1129 extension and the HH~1129-Jet in Figure~\ref{fig:SEcolor}. This sample was divided along 
lines pointing toward PA~=~336\degr\ (parallel to the two targeted features) with sample spacing of the spectra of 0\farcs5. The resulting
profiles (increasing distance from the eastern edge of the sample) are shown in the left-hand panels of Figure~\ref{fig:geheimnis}.  

Scanning with increasing displacement we see that there are density peaks associated with the passage of the the HH~1129-Jet (1\farcs0) and 
HH~1129 (1\farcs7. Much, but not all of the \Te\ increase at 1\farcs0 and 1\farcs7 can be attributed to the local rise of \Te. The spatial resolution of the 
spectra preclude drawing detailed conclusions from them. However, it is clear that \vrad(\nii) drops upon entering the region to the west and 
the FWHM(\nii) increases. Values of \Vblue\ drop after reaching the HH~1129 boundary. 

The low displacement spectra are interesting in that the \vrad\ values of about 29 \kms\ are larger than the assumed velocity (27.4$\pm$1.5 \kms)  of the Orion Molecular Cloud (Section~\ref{sec:Vcorr}. This argues that the radial velocity of the PDR does change slightly across the \hr\ or that this particular region 
(east of our sample) has a collective motion into the host molecular cloud. The spectra are also unusual in that the S(blue)/S(mif) values are much
stronger than elsewhere (the next highest is 0.18 in the Secondary Bright Bar) reaching a peak of 0.38, then dropping to being too faint to detect. As in other sections, the blue component velocities are stronger when the velocity is more positive.  In addition, there is no detectable backscattered redshifted component for any of
these samples. This is consistent with the idea that we should not expect to see a backscattered component 
when viewing a region tilted edge-on.

The radial velocity changes indicate that one progresses from a highly tilted region into a MIF that is flatter, with the increase of local \Ne\ and \Te\ occurring at the position of this transition. The HH~1129-Jet is of higher density than the MIF in that direction and probably much higher since the
object is intrinsically small. There is no dip in S(\oiii) at its position, indicating that it is not in a strongly emitting \oiii\ zone. 

\section{Discussion}
\label{discussion}
The wealth of information derived in the subsections of Section~\ref{sec:overview} has allowed us to examine multiple facets of the structure and
physics of the \hr. The first is the unexplained line broadening found in high velocity resolution studies of the \hr\ (Section~\ref{sec:fwhm}). Next
we were able to establish that the variations of \Te\ are determined by variations in \Ne, down to a scale of a few arcseconds (Section~\ref{sec:NeTe}).
Finally, we determined that the lack of a red-shifted back scattered component is due to viewing highly tilted regions of the MIF (Section~\ref{sec:BackScat}).

\subsection{Unexpected large values of the FWHM}
\label{sec:fwhm}

It has been recognized for several decades that the FWHM of lines in the \hr\ are unexpectedly high.
This was initially recognized in the photographic study of \citet{wil59} and was interpreted by \citet{mun58} as fine-scale turbulence. Typically this unexplained
component is equal to or larger than the thermal broadening of the lines, which means that we are not able to explain a large amount of energy carried in the lines. We address this problem in this
section, first summarizing earlier studies (Section~\ref{sec:previous}, then presenting and testing a simple model for explaining the extra broadening (Section~\ref{sec:FWHMmodel}). Finally (Section~\ref{sec:munch}) we summarize the results of the discovery study and discuss their relevance today.

\subsubsection{Previous studies of the unexplained component of the FWHM}
\label{sec:previous}

A slab of ionized gas will produce emission-lines of a finite FWHM. In the simplest case, the observed FWHM will be the quadratic sum
of  the thermal broadening component (5.44 \kms\ for \nii\ at 9000 K) and the instrumental broadening component (typically 8 to 10 \kms).
If there is an additional broadening mechanism that is random, it too will add quadratically; but, if it is small and systematic, it will add arithmetically.  Here we will designate the additional components as the 'Extra Line-Broadening Component' (ELBC). 

The highest spectral resolution study of the ELBC is that of \citet{ode03} , where spectra at 6 \kms\ resolution were used to study a group of 10 slit spectra, nine of which were inside the \bb\ and one outside. They found the quadratically extracted extra component for multiple emission lines:  \nii\  658.3 nm, 10.6$\pm$1.4 \kms; \Ha, 20.9$\pm$1.3 \kms; \Hb, 18.8$\pm$2.2 \kms; \Heplus, 587.6 nm, 18.4$\pm$1.9 \kms; \oi , 630.0 nm 9.0$\pm$2.1 \kms; \oiii, 500.7 nm, 13.0$\pm$3.7 \kms; \oiii, 495.9 nm, 11.3$\pm$2.1 \kms; \sii, 673.1 nm, 11.3$\pm$2.4 \kms; \siii, 631.2 nm, 11.8$\pm$1.9 \kms. They also report the radio H65$\alpha$ value of 19.6$\pm$0.9 \kms \citep{wil97}, and Jone's (1992) value for \oii\ of 10.5$\pm$2.5 \kms. \citet{gar08}
demonstrated that the \Ha\ and \Hb\ lines need to be corrected for fine-structure. After correction for this, ELBC(\Ha )~=~19.5 \kms\ and 
ELBC(\Hb )~=~18.0 \kms. 

The quadratically extracted ELBC values can be compared and grouped according to the common regions of origin and their dependence on \Te : Group 1 (\sii, \nii, \oii, \siii) has an average of 11.1$\pm$0.6 \kms; Group 2 (\oiii) 12.2$\pm$2.5 \kms, and  Group 3 (\Ha, \Hb) 18.6$\pm$0.9 \kms). 
The H65$\alpha$ value was not included in Group 3 because the large beam width of the radio observations means that large-scale motions must 
contribute to the observed FWHM.
The \oi\ emission (ELBC = 9.0$\pm$2.1 \kms) does not
fall into any of these groupings since it arises immediately at the MIF. Group 1 originates in the \nzone,  Group 2 in the \ozone, and Group 3 in both the \nzone\ and the \ozone\ and its emissivity rises with lower \Te, while the emissivity in the collisionally excited forbidden lines increases with \Te.

\citet{ode03} pursued
an explanation that the ELBC in hydrogen (they included the similar numbers for \Heplus) and the other ions was due to the fact that the recombination lines (hydrogen and helium) arise from the cool gas along the line-of-sight, whereas the collisionally excited forbidden lines arose in the hot component. When grouped according to the zones of emission, this interpretation continues to make sense because the 
Group 1 and Group 2 values are very similar, 11.1$\pm$0.6 \kms and 12.2$\pm$2.5 \kms\ respectively, with only Group 3 being different (18.6$\pm$1.5 \kms).

We should note that \cite{gar08} analyzed the widths of many lines in the Spectral Atlas by an alternative line profile fitting method and a corresponding statistical treatment. Their values of the FWHM yield ELBC components a few \kms\  greater in value and somewhat greater uncertainty for the forbidden lines (14.3$\pm$3.9 \kms) than in the \citet{ode03} study (11.2$\pm$2.3 \kms). Their ELBC for \Ha\ is 18.6$\pm$5.1 \kms, compared with our Group 3 value of 18.6$\pm$1.5 \kms. Through a detailed argument used in deriving \Te\ from the FWHM values for several
ions, they conclude that there is no difference between the ELBC for collisionally 
excited and recombination lines. The complexity of their approach creates some uncertainty in this conclusion and we will consider in our discussion both possibilities, i.e. that the ELBC for cool and hot gas is the same or different . In any event, the ELBC as measured by collisionally excited lines is indistinguishably the same in the two ionization zones present in the \hr.

The advantage of our study is that we have considered separately the results from many samples for the \nii\ 658.3 nm using the Spectral Atlas.   Although \citep{gar08} states that the instrumental resolution is 10 \kms, the data is from several sources with a range of resolutions. The spectra of \citet{doi04}, have a resolution of 8 \kms\ and were used for the
\nii\ and \oiii\ lines employed in this study. It is possible that the resolution was degraded as the original spectra were cast into the convenient form of the Spectral Atlas. For this analysis we adopt 10 \kms. 

In addition to \vrad, the determination of the FWHM is an integral product of using the task ''splot''. When a line gave even a hint of being multiple (through an unusually high value of FWHM
or an asymmetry of the line profile, an attempt was made to fit multiple components. This was only necessary on rare occasions. The accuracy of \vrad\ is usually better than 1 \kms\ and the 
determination of FWHM was usually better than 2 \kms.

\subsubsection{A role of photo-evaporative flow in determining the FWHM?}
\label{sec:FWHMmodel}

\begin{figure}
	\includegraphics[width=\columnwidth]{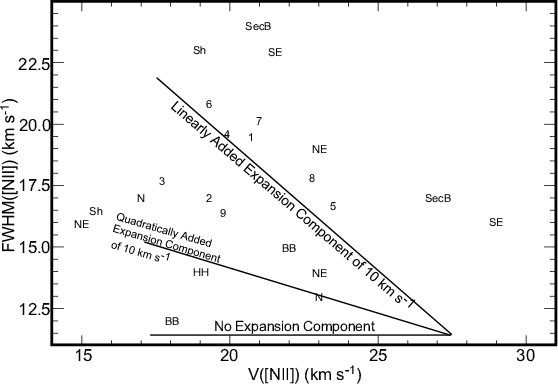}
    \caption{ 
    This figure compares the observed values of \vrad(\nii) and FWHM(\nii) with the 
    predictions of the simple model presented in Section~\ref{sec:FWHMmodel} where the extra broadening is due to photo-evaporative flow. The lower line is the expected relation if there is no effect of photo-evaporative flow, the second line applies if the component is added quadratically,  and the upper is the expected relation if photo-evaporative flow is linearly added to the FWHM. The numbered samples are those
    from the West Column (Section~\ref{sec:NorthernArea}), Sh is for the Shadow Sample (Section~\ref{sec:shadow}),
    N for the North Sample (Section~\ref{sec:north}), SE for the Southeast Sample (Section~\ref{sec:SE}),
    HH for the HH~204 Sample (Section~\ref{sec:HH204}), NE for the Northeast Arc (Section~\ref{sec:NE}),
    BB for the Bright Bar (Section~\ref{sec:BrightBar}), and SecB for the Secondary Bright Bar Sample (Section~\ref{sec:narrow}).
    }
     \label{fig:elbc}
\end{figure}

The overall blue-shifting of lines from different stages of increasing ionization gave rise
to the widely accepted blister model of the \hr\ and in this paper we have established that the local variations in \vrad\ can be explained by 
variations in the tilt of the ionization front. Evaporative expansion certainly occurs and it has previously been argued from models of the photo-evaporative flow that it cannot 
account for the ELBC \citep{hen05a}. Never-the-less, we found it useful to test that conclusion using both FWHM(\nii) and \vrad(\nii) data together in a comparison with a simple model.

There are two ways of possibly adding the effects of photo-evaporating flow to what we observe. The results of these two approaches are shown 
in Figure~\ref{fig:elbc}.
If the FWHM is determined by quadratic addition of the thermal (5.44 \kms\ for \Te= 9000 K) and instrumental broadening (10 \kms), then the expected
FWHM would be constant at 11.4 \kms. The photo-evaporating flow accelerates from the velocity of the underlying PDR towards the observer reaching
a velocity of about 10 \kms\ averaged over the \nii\ emitting layer. When one views the MIF where it is tilted edge-on, the photo-evaporating
flow will not contribute to the FWHM and the observed \vrad(\nii) will be that of the PDR. This would be a single value of \vrad(\nii) =27.4 and FWHM(\nii) = 11.4. In the case when one views the MIF flat-on, one expects \vrad(\nii)= 17.4 \kms\ and the FWHM will be increased because of the range of velocities
within the \nii\ emitting layer. If this component added quadratically, the FWHM at \vrad(\nii)= 17.4 \kms\ would increase to 15.2 \kms. However,
this broadening component is systematic, rather than random, and an arithmetic addition is a more accurate description. In this case the 
value at \vrad(\nii)= 17.4 \kms\ would be 21.4 \kms, although such an extreme result is unrealistic because in that case 
the method of using "splot" would identify such a line as double, rather than producing a wide single-line component. This means that the 
upper line in Figure~\ref{fig:elbc} is unrealistically high.

The only certain conclusion that can be drawn from Figure~\ref{fig:elbc} is that all of the data points fall on the same plane!
Although photo-evaporative flow must play a role in producing the non-thermal line broadening, it can only be inferred for the points
lying well beneath the upper line.  The effect must be present, but is not established from this data-set. Certainly the points lying above
the upper line are not produced in this way. This would be true even if one used the \oiii\ expansion velocity (which would certainly
apply to the ionized hydrogen emission) of 15 \kms.

These conclusions mean that although photo-evaporation must play a role in producing the anomalous FWHM, there remains 
an unexplained major contributor.  The study of \citet{ode03} remains an important guide and if the ELBC for ionized hydrogen is
much larger than for the collisionally excited ions, then line-of-sight variations in \Te\ is the important factor, rather than turbulence.
However, if there is not a systematic difference in the ELBC then turbulence is probably the source. In fact, the simulations of a photo-evaporative gas by \citet{mel06} show that this can be generated by flows off of dense knots within the PDR.

\subsubsection{Comparison of the results of this study with an earlier velocity study of the \hr}
\label{sec:munch}
In  a classic study of the \hr\  \citet{wil59} mapped the face of the nebula with multi-slit spectra in multiple emission lines. This work and the conclusions
that could be drawn were summarized in \citet{mun58}. The velocity and angular resolution was comparable to that in the Spectral Atlas. The detector was the photographic emulsion and the difficulty of obtaining photometric results caused the authors to simply study the radial velocities and FWHM. Figure 4 of 
\citet{mun58} shows the velocity results for an east-west sample and this resembles the nearby data presented in our Figure~\ref{fig:northern}, which is of a north-south sample.

However, their interpretation of the velocity variations as being large-scale random turbulent motions is probably incorrect. The adopted model of the nebula was an ionized gas whose dust component renders it optically thick to its own emission, making it appear to be a flat surface-which had been treated in the recent theoretical papers on turbulence. Subsequently, there
were several similar studies done by the lead author and his students  based on the blister model (c. f. Section 3.4.2 of \citet{ode01}), interpreting the velocity variations within the paradigm of large-scale turbulence. We now
know that the variations in velocity are primarily caused by the photo-evaporating ionization front being viewed from different angles along the line of sight. Moreover, recent theoretical work \citep{med14} shows that one cannot interpret the variations as turbulence. 
Never-the-less, \citet{mun58} correctly identified the source of the non-thermal FWHM as being due to turbulence.

\subsection{A Summary of \Te\ and \Ne\ variations and their causes}
\label{sec:NeTe}

The study of many samples throughout the \hr\ has given us a good idea of the magnitude of the \Ne\ and \Te\ variations, the relation of these two 
quantities, and their spatial scale. In this section we describe the relation of \Ne\ and \Te\ , compare this relation with predictions of the best model
for the \hr, then describe the contributing factors to the \Ne\ variations.

\subsubsection{Systematic Changes in  \Ne\ and \Te }
\label{sec:empirical}

\begin{figure}
	\includegraphics[width=\columnwidth]{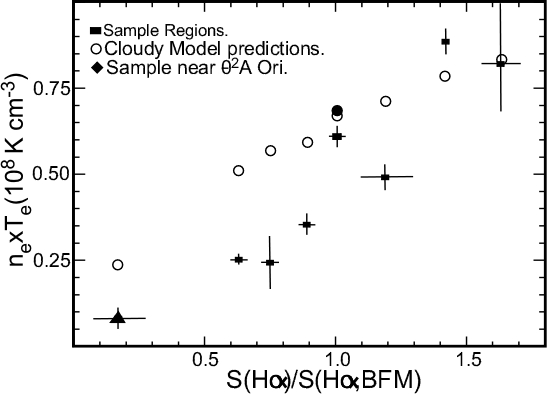}
    \caption{ 
    The \Ne$\times$\Te\ values using data in Table~\ref{tab:WestColumn} and Table~\ref{tab:NeTe} were grouped into samples of similar S(\Ha) and are shown as filled rectangles, with the S(\Ha) values taken from the MUSE extinction corrected image. The S(\Ha) values are normalized to the reference sample of \citet{bal91} (BFM).
 {\it Sample 5 is from Table~\ref{tab:WestColumn} and the open rectangle is for the Bright Bar. The reason for their horizontal error-bar arrows are discussed in the text.}
The lower-left filled triangle is a sample beyond the \bb\ that is probably partially photoionized by \ta.
The open circles are the results of models similar to that in BFM as described in the text.
The filled circle gives the \Ne$\times$\Te\ value calculated from the line ratios in BFM.
}
    
     \label{fig:NeTe}
\end{figure}

Examination of the \Ne\ and \Te\ values in the nine groups within the West Column as presented in Table~\ref{tab:WestColumn} indicated a correlation 
between these two properties and we have extended this data-set by determining these values at the peak and backgrounds for the
other samples. The ten additional results are given in Table~\ref{tab:NeTe}. In this case we cannot accurately determine the uncertainty of each entry because we
are picking points from a graph, whereas for the West Column samples the uncertainty is derived from the spread of values within the 
group. All of these results are summarized in Figure~\ref{fig:NeTe}.

\begin{table*}
\centering
\caption{Observed \Ne\ and \Te\ values}
\label{tab:NeTe}
\begin{tabular}{lcccc}
\hline
Region    &  Background   & Peak              & Observed                 & Observed\\
                & \Ne~(\cmq)     & \Ne~(\cmq)    & Background \Te~(K) & Peak \Te~(K)\\
 \hline
 HH~204 &    1000            & 5000              & 8000                          & 9300\\
 Bright Bar & 1800            & 3400              & 8750                         & 9100\\
 Sec.Brt.Bar & 2700          & 4700             & 9200                         & 9800\\ 
 Shadow &  2850              & ----                 & 8900                         & ----\\
 NorthSample&5500          & ----                 & 10500                        & ----\\
 NE Arc &      3000             & 6500             & 9700                          & 10900\\
\hline
\end{tabular}
\end{table*}


\subsubsection{The predicted relation of \Ne$\times$\Te\ and S(\Ha) }
\label{sec:prediction}


The observations presented in Figure~\ref{fig:NeTe} demonstrate a correlation between S(\Ha) and  \Ne$\times$\Te. The former quantity is proportional to the flux of ionizing photons and the latter quantity 
is proportional to the gas pressure near the MIF as derived from \sii\ and \nii\ emission lines. A correlation of these quantities is expected for a blister model of an \hii\ region in equilibrium.

In such a model, the geometry consists of an inner bubble of very hot (T about 10$^{7}$ K) stellar-wind-shocked gas, with the \hii\ forming a boundary between this bubble and the surrounding molecular cloud.  The \hii\ region pressure is established by the combination of radiation pressure in the absorbed stellar radiation field, which often dominates, and the pressure in the inner hot bubble. 

Predictions of the best extant model for the \hr\ are also shown in Figure~\ref{fig:NeTe}. These Cloudy models 
 were built upon the successful BFM model of an inner region of the nebula using the Cloudy package \citep{gjf13}.  We assumed that S(\Ha) indicates the conditions at the illuminated face of the \hii\ region. We took S(\Ha) at each position as an initial condition and then computed a model in which all the forces are in balance.This tests the hypothesis that radiation pressure sets the density.
 
 The predicted values of \Ne$\times$\Te\ and S(\Ha) agree with the observations in the brightest part of the nebula (except for Sample 5 which is discussed in the following section). However, at lower surface brightnesses the observed \Ne$\times$\Te\ is lower than predicted.  

\subsubsection{Comparison of the Observed and Predicted values of \Ne$\times$\Te }
\label{sec:prediction}

The observed points in Figure~\ref{fig:NeTe} are subject to systematic effects. When the local samples are highly tilted the   \Ne$\times$\Te\  values will be unchanged since this depends upon the ratio of lines within a narrow ionization zone. However, S(\Ha) 
will be enhanced due to the projection effect along the tilted MIF. We see this effect most dramatically in the Bright Bar sample,
which is known to be highly tilted and in Sample 5, where \vrad\ reaches 26.8 \kms\ indicating a highly tilted region. The amount
of the overestimate of S(\Ha) is uncertain, but would be two times in the case of a tilt of 60\degr. This means that the Bright Bar and Sample 5 values of S(\Ha) must be corrected by an unknown but large amount to lower values.

The fact that the lower S(\Ha) values systematically lie lower than the predictions of the models can be explained by the 
fact that the ratio of radiation pressure at the MIF and the hot core gas has been assumed constant. However, this need not
be the case as the radiation pressure will drop inversely as the square of the distance from \tc\ and the hot core pressure
will decrease more slowly. 

Our models have not included any pressure term for turbulent pressure, that is, the random motion of packets of gas. 
Inclusion of such a pressure term would drop the predicted value of  \Ne$\times$\Te\  by an amount depending upon 
the density and turbulent velocity. If this is the cause of the disagreement in the low S(\Ha) samples, then this indicates
that turbulence becomes relatively more important there. One of the important conclusions of this paper is the confirmation
of the conclusions of \citet{ode03} that there is fine-scale structure at the MIF. The local flow off of these volume elements will 
simulate turbulence.

\subsubsection{The multiple features that produce variations in \Ne}
\label{sec:NeComponents}

One component of the variations in \Ne\ are variations in densities within the PDR, which in turn reflect variations
in density of the nearby molecular cloud. Photo-evaporation of gas from the PDR is a very dynamic process, with material being
removed. The ionization front will advance into the molecular cloud in those regions of lowest density, thus isolating the higher
density knots. On a large-scale this is probably what produces the structured surface of the ionized layer, with the most extreme
example being the dense Orion-South cloud. In the case of the Orion-South cloud the ionization front has eaten its way behind the cloud,
so that we see in optical wavelengths radiation from the near side of the cloud and at radio wavelengths emission from both the              f
front and rear. Molecular and 21-cm absorption lines indicate that the Orion-South cloud has a 
neutral molecular core \citep{ode10,vdw13}. A second component of the variations arises from when the MIF is tilted to face \tc, the 
flux of ionizing photons increases and the rate of photo-evaporation increases, producing a local rise in density. We see this in 
the tilted features like the Bright Bar having higher \Ne\ values than the local region.

The hypothesis of there being significant density variations in the PDR and host cloud is also supported by the fact that the
radial velocity changes across the nebula. The clearest example is in the West Column samples, where \vrad(\nii) changes
significantly over size scales of 10\arcsec\ or less. Since these variations are caused by varying tilts in the ionization front, the
velocity variations mimic the fine-scale structure.  
There is corroborating evidence for clumpy structure behind the PDR in the high resolution radio studies of molecules in the
Bright Bar \citep{vdw96}. 

The result that there are variations of \Ne\ and \Te\ down to the limits of resolution of this study may have a role in explaining 
the Abundance Discrepancy Factor (ADF), the fact that one derives different relative abundances for heavy elements when using recombination or collisionally excited forbidden lines \citep{pei67,pc, mesa08}. The ADF was not resolved by proposing the existence of a non-thermalized
electron energy distribution (commonly called a Kappa distribution) as shown in \citet{fer16}. It is most likely that the explanation lies
in the small scale \Ne\ and \Te\ variations that also give rise to the turbulence component of the line broadening.

This study has mapped the \Te\ variations in two dimensions, with single samples being an integral along those 
lines-of-sight. If small-scale density structures exist, then those structures will have their \Te\ increased because
of suppression of forbidden-line heating, but also through increased dissipation of turbulence and shocks.

In this study we have been limited by the sample size necessary to obtain a high S/N ratio. A more discriminating study would require 
a spectrophotometric map at HST imaging resolution, accompanied by matching high velocity resolution spectroscopy.  The former
goal is being pursued by W. J. Henney.

\subsection{Samples without a Backscattered Component}
\label{sec:BackScat}

A red component in the \nii\ emission-line is present in almost all of our samples. 
This is to be expected within the paradigm that it is backscattering from dust in the underlying PDR.
Within this model, one would expect it to be most visible in sections of the MIF seen face-on
and to be weak or absent in those seen edge-on.
This means that the fact that it is absent must be telling us about the local conditions.

It is absent in all slits of the SE sample. The \vrad\ is about 29 \kms\ in the lower displacements and 22 \kms\ in the higher. 
The appearance of image of this region is that of an open separation between two volumes, with the HH~1129-Jet passing between. The high \vrad\ of the smaller displacements 
indicates a highly tilted region and the lack of backscattering is to be expected.  The absence at higher displacements is puzzling, but this is an admittedly complex region.

Backscattering is absent in slits 1--4 of the Shadow, where \vrad\ is higher (19 \kms), than the remaining region (15.5 \kms). Emission from the edge of a low ionization column
within the \ozone\ layer should not have backscattering, which is probably the interpretation of this object.

There is no backscattering in the Secondary Bright Bar  spectra. It is a situation similar to that of the SE sample, where one side is very high \vrad\ (27 \kms) , hence highly tilted.
However, we again have a situation where the nearby regions of lower velocity (21 \kms) share the lack of a backscattered component.

In the West Column samples the red component is absent at lines 11--14, 41--50, and 63--76. These appear at displacements about 2\farcs5 greater than the line numbers. 
These coincide with the three \vrad\ maxima in the West Column (24 \kms, 27 \kms\ and 24 \kms). 

Except for the low displacement Shadow samples (where there is an alternate explanation), all of the samples without backscattering are associated with a highly tilted region of the MIF.
At the same time, there are no high velocity samples that do have backscattering, thus strengthening the conclusion that backscattering arises from the PDR and is not seen when
the PDR is highly tilted.

\section{Conclusions}
\label{sec:conclusions}

The variations of radial velocity of the MIF components can be explained as viewing a photo-evaporating ionization front from various angles relative to the line-of-sight.

\Te\ and \Ne\ variations are common across the \hr, and there is a 
patten of \Ne$\times$\Te\  varying with the local surface brightness in \Ha. Comparison with models of the ionized layer indicates 
good agreement for the high surface brightness samples. The systematic differences in the low surface brightness samples
 indicates that the assumptions of the model must not be exact there.

There is a blue shifted spectral component in almost all of the regions studied and this often varies in velocity and strength in the same places that the MIF velocities vary.

The absence of a red shifted backscattered component is well associated with regions of high \vrad\ that must be regions viewed more edge-on.

All of the samples show FWHM higher than expected from thermal and instrumental broadening. Large-scale photo-evaporative flow must contribute at some level, but certainly cannot explain this excess. It is necessary to invoke a broadening component originating from fine-scale knots present at the ionization front and into the \nii\ emitting zone. Photo-evaporation from these knots probably drive turbulence in the ionized gas.

  An attempt to determine \Te\ for \Hplus\ from combined radio and optical observations showed a general agreement with \Te (\nii), but more
important, that the wavelength dependence of the extinction transitions from the more grey extinction in the \hr\ to the colour dependence of the general
interstellar medium at the position of the Bright Bar.

\section*{Acknowledgements}

We are grateful to Wolfram Kollatschny, Peter Weilbacher, and the remainder of the MUSE team for creating the MUSE
data-base of \hr\ spectra and to William J. Henney for the processed GO-12543 data. Partial support for CRO's work was provided by HST grant GO 12543, whose original
principal investigator was the late Robert H. Rubin. GJF acknowledges support by NSF (1108928, 1109061, and 1412155), NASA (10-ATP10-0053, 10-ADAP10-0073, and ATP13-0153), and STScI (HST-AR- 13245, GO-12560, HST-GO-12309, GO-13310.002-A, HST-AR-13914, and HST-AR-14286.001). MP received partial support from CONACYT grant 241732.

 This paper benefitted in content and presentation from the comments of the referee.
 







\appendix

\section{{\it An attempt to determine \Te\ for \Hplus\ yields information about the extinction curve}}
\label{sec:IntroExt}



The common assumption in determining relative abundances is that the \Te\ found from forbidden lines also applies to the coexisting \Hplus. If this is not true, it will affect the derived relative abundances. In the following section we describe previous methods 
used to independently derive the \Te\ that applies to \Hplus\ and in the following sections describe an attempt using a new method.

\subsection{Earlier Methods for Determining \Te\ for \Hplus}
\label{sec:OldTeMethods}
In a pure hydrogen gas one can derive the electron temperature of \Hplus\ from the relative strength of the \Hb\ line to the underlying continuum (AGN3), but in the 
case of the \hr, there is a strong stellar scattered light component that is much stronger than the nebular emission \citep{bal91,ode10}. This dominant 
contaminating component prevents one from determining accurate values of \Te\ by the optical line-continuum ratio method. 

\Te\ can also be determined 
from the ratio of surface brightness in hydrogen radio recombination lines and the underlying continuum (which has no stellar scattered light component).
The studies by \citet{wil97} and Wilson, Bania $\&$ Balser (2015), illustrate the power of this method, but the state of the art is that the angular resolution is about
2\farcm5\, which covers the \hr. For reference the distance between the brightest star in the Trapezium (\tc) and \ta\ (which lies
to the southeast and just beyond the \bb\ feature) is 2\farcm24. The \citet{wil15} study found \Te\ = 8300$\pm$200 K, which is consistent, after consideration of their lower spatial resolution, with the \Te\ values that we find.

\subsection{A new method for determining \Te\ using radio continuum and optical lines}
\label{sec:introNew}

Comparison of the surface brightness of the nebula in the radio continuum with a Balmer emission line can be used to determine
the extinction.  This was first explored by \cite{ode00b}. They used Very Large Array (VLA) observations of the free-free continuum of \Hplus\  and an assumed constant \Te\ to predict the surface brightness in the Balmer \Ha\ line.  The ratio of the observed and predicted \Ha\ surface
brightness (\SHa) gave an indication of the amount of extinction, expressed as \cHb (the logarithmic extinction at \Hb), defined as  
\cHb\ =  log\ (\SHbObs /S(H$\beta$,Predicted). This result is dependent upon the assumed \Te.

The more common method of establishing the amount of extinction is from measurement of the observed \Ha/Hb\ ratio and 
interpreting the difference from an assumed intrinsic value as due interstellar reddening, for example \citet{wei15}. The assumed intrinsic 
line ratio is dependent on the assumed \Te. 

Since both of these two methods assume a value of \Te , it is possible to eliminate \cHb\ as a variable and derive \Te\ 
for the \Hplus\ emission.


\subsubsection{Derivation of \cHb\ from radio and optical observations} 
\label{sec:Cvla}

In their study of the extinction \citet{ode00b} employed observations made with the Very Large Array (VLA) together with calibrated images of the \hr\ made with the Hubble Space Telescope (HST). The VLA observations of the continuum were made at 20.46 cm wavelength and have a FWHM resolution
of 1\farcs83 by 1\farcs65, with the long axis pointed towards PA=-34 \degr. 
This image was converted to units of \sbunits.
Because the HST images were more than an order of magnitude higher resolution, 
they were degraded using the IRAF task ``gauss'' to match the VLA image. The VLA image was used to predict the expected surface
brightness in the \Ha\ line. A comparison of the observed and predicted surface brightness and the assumption of the form of the extinction law allowed 
the derivation of \cHb\ at this resolution across the \hr.  They then qualitatively compared the resulting map of \cHb\ with that derived from the \SHaObs /\SHbObs\ 
ratio over a small portion of the \hr. 
In a later paper \citep{ode09b} these results were discussed further and it was shown in their \S 2.4 that the interpretation of the extinctions derived by the two methods was more complex. 

In the present study we employ the earlier VLA observations, but utilize the \SHbObs\ images derived at about 1\arcsec\ resolution (sampled at 0\farcs2) during 
the commissioning of the MUSE spectrograph \citep{wei15}. For the remainder of this paper we will designate as \Cvla\ the \cHb\ value derived from a combination of VLA and \SHbObs\ and the \cHb\ value derived from the \SHaObs/\SHbObs\ ratio as \Cmuse.

\citet{ode00b} describe the methodology of deriving \cHb\ using the VLA image using the formulation in an earlier version of AGN3. Our approach is the 
same, but we express the mathematical relations using \SHbObs\ in \sbunits rather than photons~s$^{-1}$~cm$^{-2}$~sr$^{-1}$. More important is that we now include the role of variations in \Te, which was assumed to be constant (9200 K) by  \citet{ode00b}.  

The surface brightness in the radio continuum will be

\begin{equation}
\rm I_{\nu}=\frac{2kT_{e}}{\lambda^{2}}(1-e^{-\tau_{\nu}})
\label{eq:radio}
\end{equation}
ergs cm$^{-2}$ s$^{-1}$ hz$^{-1}$ ster$^{-1}$, 
where k is Boltzmann's constant, $\lambda$ the wavelength, and $\tau_{\nu}$ the optical
depth in the radio continuum through free-free processes at $\lambda$.
\SHb\ will be
 \begin{equation}
\rm S_{H\beta}=\frac{3.086x10^{18}}{4\pi}\alpha^{eff}_{H\beta}E_{H\beta}EM
\label{eq:SHb}
\end{equation}
ergs cm$^{-2}$ s$^{-1}$ ster$^{-1}$,
where $\rm \alpha^{eff}_{H\beta}$ is the emissivity of H$^{+}$, $\rm E_{H\beta}$ is the 
energy in an \Hb\ photon, and EM is the commonly used term ``Emission Measure'' (the integral of \Ne$\times$N(\Hplus) along the line of sight) in units of cm$^{-6}$pc. \Ne\  and N(\Hplus) are the densities (\cmq) of electrons and ionized hydrogen respectively.

$\tau_{\nu}$ is also proportional to EM.  AGN3 gives 
\begin{equation}
\rm \tau_{\nu}=0.0824 T\rm_{e}^{-1.35}\nu^{-2.1}_{ghz} EM. 
\label{eq:tau}
\end{equation}
There will also be an absorption 
component from singly ionized helium (the highest state of ionization of helium found in the \hr), so that $\tau_{\nu}$ is increased by about 10$\%$.

Introducing the common notation that t = \Te/10000 and combining equations \ref{eq:radio}, \ref{eq:SHb}, and \ref{eq:tau} gives
\begin{equation}
\rm c_{VLA} = log~\Bigl\{ \frac{t^{1.35}f_{eff}[-~ln(1-I_\nu /t \times 6.596 \times 10^{-15})]}{5.32\times S(\rm H\beta,OBS)}       \Bigr\}
\label{eq:Cvla}
\end{equation}
where f$\rm _{eff}$ is the electron temperature dependent scaling factor for $\alpha^{eff}_{H\beta}$.
Using the values in Table 4.2 of AGN3, we find f$\rm _{eff}$ = t$^{-0.83}$. 
The resulting \cHb\ map of the \hr\ for \Te~9200 K is very similar to Figure~2 of \citet{ode00b} 

\subsubsection{Derivation of \cHb\ from the \SHaObs\ to \SHbObs\ ratio}
\label{sec:Cmuse}
The derivation of \Cmuse\ is less involved as it involves only \SHaObs\ and S(H$\beta$,OBS).
It is calculated from 
\begin{equation}
\rm c_{MUSE} = \frac{1}{\bf{r}}\Bigl\{ log \frac{S(H\alpha,OBS)}{S(H\beta,OBS)}  - log\Bigl[\frac{S(H\alpha)}{S(H\beta)}\Bigr]_{t=1} - log~ g(t)   \Bigr\}
\label{eq:Cmuse}
\end{equation}
where [\SHa / \SHb]$\rm_{t=1}$ is the line ratio at \Te = 10000  (2.87 from AGN3) and g(t) is a 
function of the electron temperature and is approximately g(t)=t$^{-0.086}$ (again using the numbers in Table 4.2 of AGN3). 

The greatest uncertainty in this derivation is in the wavelength dependent extinction factor. It is well known that the Orion reddening curve
is much flatter (more like grey extinction) than that for the general interstellar medium (ISM). The most thorough investigation is that of \citet{bla07}
where the shape of the Orion Extinction Curve was derived most accurately. They express the Extinction Curve, f$_{\lambda}$, normalized to zero
at the wavelength of \Hb. In their nomenclature {\bf r} = f$_{486.1\ nm}$-f$_{656.3\ nm}$, which is 0.22 as calculated from their Table 2. {\bf r} would be much larger
using the Extinction Curve from the ISM and would be smaller for even more grey extinction. Our derived \Cmuse\ map is very similar to that of \citet{wei15}, where they assumed an intrinsic and flux ratio of 2.89 of \Ha\ and \Hb. 

\begin{figure*}
	\includegraphics
	[width=7.5in]
	{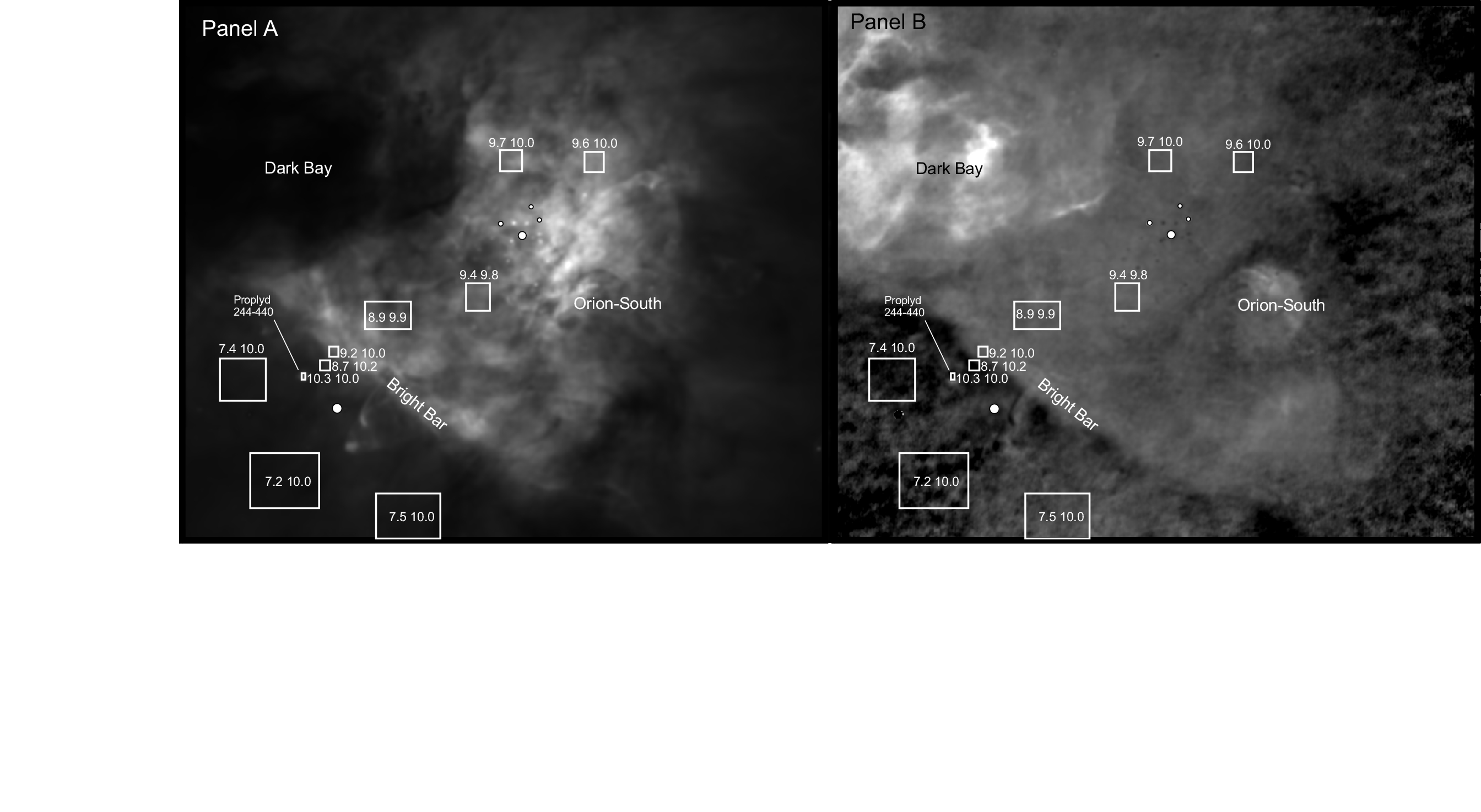}
    \caption{
This pair of of 353\arcsec $\times$295\arcsec\ images of the \hr\ shows the full FOV image of the  MUSE database \citep{wei15} in \Ha\ in Panel A and the \Cvla/\Cmuse\ results described in Section \ref{sec:introNew} in Panel B.
    North is up and east to the left.
    The brightest four of the Trapezium stars are shown as filled circles as is \ta\ in the southeast.
    The \Ha\ image has been degraded to 1\farcs83 $\times$ 1\farcs65 FWHM and both panels have a scale of 0\farcs5 per pixel. 
    The \Cvla/\Cmuse\ results were calculated with a fixed temperature of 10000 K. 
    The sample regions subjected to a detailed analysis
    Section \ref{sec:cratio} are shown as boxes. 
    Within or near the boxes the left-hand number is the temperature derived from the MUSE \nii\  auroral and nebular lines in units of 1000 K while the right-hand number is the temperature, in the same units, derived for the \Hplus\ emission, 
   using the \Cvla / \Cmuse\  results. 
   The sample centred on the bright portion of proplyd 244-440 is also shown, to the northeast of \ta.
   }
     \label{fig:extinction}
\end{figure*}

\subsection{Using \Cvla /\Cmuse\ as a temperature indicator}
\label{sec:cratio}
The intrinsic ratio of \Cvla\ and \Cmuse\ is dependent on \Te. This means that this ratio can be used to derive \Te\ for the \Hplus\ 
emitting regions, rather than assuming \Te\ to be the same as that indicated by co-existing collisionally excited forbidden lines. 
The numerator in the \Cvla /\Cmuse\ ratio is the more sensitive to temperature.
There is a log (t$\rm ^{1.35}$ $\times$ f$\rm_{eff}$) term in the numerator (this is about 0.52$\times$log t) and a t$\rm ^{-1}$ factor in the I$_{\nu}$ term (both in equation \ref{eq:Cvla}). The denominator is less sensitive to temperature as it determines the log g(t) term in equation \ref{eq:Cmuse}, which is about -0.086$\times$log~t.

The right-hand panel of Figure \ref{fig:extinction}  shows a map of \Cvla /\Cmuse\ calculated with t~=~1, while the left-hand panel presents a MUSE \Ha\ image degraded to the VLA resolution.  The centre of the \hr\ is about 0.8 and other regions and features are greater or lesser than this. 

The values in the Dark Bay and Orion-South region are larger than 1 (reaching a peak of 2.4 in the Dark Bay and 1.2 in Orion-South). \citet{ode00b} noted that these regions are where the radio and Balmer line emission don't originate from the same volumes. 
In these areas the extinction arises from in front of the volumes producing the optical lines but there is background optical emission
that is obscured by an intervening extinction layer. In the case of the Dark Bay the extinction is formed from a dense
area within the foreground Veil and is sufficiently optically thick that we don't see optical emission from near the main ionization front of this deformed-blister nebula. In the Orion-South area we see optical emission from the near side of the optically thick Orion-South cloud and radio emission from both there and the portion of the main ionization front that lies beyond it. 

The interpretations of the \Cvla /\Cmuse\ values for the other regions are more complex. For each spot in the map equation \ref{eq:Cvla}  and equation
\ref{eq:Cmuse} have a unique value of \Te\ that matches the observed \Cvla /\Cmuse\ ratio. In general, the ratio drops with increasing temperature, but, there is no uniform absolute value for a given temperature. The temperature must be derived pixel by pixel across the \hr. We call the temperature derived from
the \Hplus\ observations \Th. 

We divided the \hr\ into 9 samples as shown in Figure \ref{fig:extinction}. These samples were selected to represent different types of regions in the nebula. 
For each sample the averaged \SHa, \SHb, and I$_{\nu}$ observations were adopted to calculate the variation of \Cvla /\Cmuse\ for a range of values of \Te. When the calculated ratio matched the 
observed ratio, that temperature was adopted and is shown in the panels. This method effectively solves for \Te, independent of the amount of extinction. The scatter in \Cvla /\Cmuse\ varied from about 5$\%$ to 20$\%$. This scatter led to uncertainties in the derived temperatures of 200 K to 1500 K. A tenth sample, centred on the largest proplyd in the \hr, was also studied in this fashion and is discussed below.
Figure \ref{fig:extinction} also shows the averaged \Te\ values derived from the MUSE \nii\ lines, which we will  call \Tn. We see that the \Tn\ values are usually smaller than the \Th\ values. 
There are systematic differences between the inner and outer parts of the \hr, where these areas are closer to the Trapezium than the \bb\ or further, respectively. For the five samples in the inner region the average ratio of \Tn/\Th\ is 0.94$\pm$0.03 and for the outer four samples it is 0.74$\pm$0.02. 

\subsection{The extinction curve appears to be different inside and outside the \bb}
\label{sec:Difr}
The \Hplus\ emission should arise from both the \nzone\ and the \ozone\ layers and photoionization models predict only small differences of \Tn\ and \Th, with \Tn\ about 10\%\ higher.
However, the observed \Tn/\Th\ ratio is less than unity and is different inside and outside of the \bb. The answer to this riddle probably does not lie in a problem with the basic blister model of the nebula and our ability to calculate \Te\ in different ionization zones, rather it must lie in the assumptions we have made 
about the nature of the extinction curve, which determines {\bf r}. Recall that for each sample region the calculated \Cvla /\Cmuse\ ratios decreased with 
increasing assumed temperature. This means that if the ratio derived from the observations is too low, then the derived \Th\ value is too high. 

In order to derive \Th\ values that are no greater than the \Tn\ temperatures, we need to look for causes of a non-representative value of \Cvla /\Cmuse. 
The quantity {\bf r} occurs in the numerator in the derivation of \Cvla /\Cmuse\ and we have used the \citet{bla07} value of 0.22 throughout. Adjusting
the derived \Cvla /\Cmuse\ to give lower temperatures means using a larger value of {\bf r}. To match the \Cvla /\Cmuse\ to that expected from \Tn\ would mean
using {\bf r} = 0.23$\pm$0.02 for the Inner samples and 0.36$\pm$0.07 for the outer samples. 

It appears that the extinction curve is very different for the inner and outer samples. The value of 0.23$\pm$0.02 is well within the apparent uncertainty
of the extinction curve as shown in Figure 3 of \citet{bla07}. Therefore it appears that the \Cvla /\Cmuse\ ratio support the unusual extinction curve in the \hr\ as determined by \citet{bla07}.  Our {\bf r} value for the outer samples agrees well with the extinction curve difference of 0.35 for the ISM presented in AGN3. 

This leads us to the important conclusion that in the outer samples the ISM extinction curve applies.
It is generally thought that the more grey extinction that applies to the \hr\ is a result of either destruction of the small size grains within the distribution of sizes or the building up of larger
grains by the coalescing of smaller grains. In any event, these results indicate that the nature of the interstellar particles causing the extinction in the \hr\ abruptly changes at the \bb. This property is one more fact that must be accommodated when building a more accurate model of the \hr.

There are important caveats about using the \Cvla /\Cmuse\ ratios. \citep{ode09b} pointed out that filter images (for example the WFPC2 F487N filter) measures all of the components, that from the MIF, the backscattered red component, and emission from the Blue Layer. 
The ratio of the backscattered component to the MIF component (around 0.1) will artificially enhance the observed surface brightness of \Hb\ and thus lead to a \Cvla\ result that is too small. The backscattered component is less important in determination of \Cmuse\ because it 
will only enter as the difference in the backscattering at the \Ha\ and \Hb\ lines. Therefore, a correction for backscattering would systematically increase the expected \Cvla /\Cmuse\ ratio by about 10\%\ and raise the \Cvla /\Cmuse\ derived \Th\ value. This would broaden the disagreement in the \Te (\nii|) and \Te (\Cvla /\Cmuse) results, which argues that backscattering doesn't vitiate the \Cvla /\Cmuse\ approach.
The Blue Layer emission will enhance both the emission line radio surface brightnesses, which argues that the Blue Layer contamination is not important. However, in Section~\ref{sec:HH204} we saw that outside of the \bb, the Blue Layer emission became important for \nii.
Since we do not know the density and temperature of the Blue Layer, we cannot estimate its effect on our \Th\ results.

It is noteworthy that in a recent study by \citet{mcl16} it was found that the proplyds outside of the \bb\ occupy a different
[S\sii + S\siii]/\SHb\ versus S\oii / S\oiii\ space than those lying inside that boundary. Their index was derived using extinction corrected surface brightness where the \Ha /Hb\ method was used. If the extinction curve is steeper outside the \bb, this property should be re-examined. The largest proplyd in the \hr\ is 244-440, which also lies outside the \bb, and was included in our \Cvla/\Cmuse\ study.
Not only is it the proplyd with the largest apparent size, it is clearly photoionized by \ta\ (the bright rim of the proplyd faces that star). 
In this case, the \Tn\  value (t=1.03$\pm$0.04) agrees well with the \Tn\ value (t=1.00$\pm$0.1), indicating that this proplyd is in a region where the inner-\hr\ extinction curve applies, even though its position projects to outside the \bb.



\bsp	
\label{lastpage}
\end{document}